\documentclass[
  journal=pasa,
  manuscript=research-paper, 
  year=202X,
  volume=YY,
]{cup-journal}

\usepackage{amsmath}
\usepackage{orcidlink}
\usepackage{amssymb,microtype,siunitx,booktabs,multirow}
\usepackage[normalem]{ulem}

\newcommand{\hi}{\mbox{H\,{\sc i}} } 
\newcommand{\hix}{\mbox{H\,{\sc i}}} 
\newcommand{\mgii}{\mbox{Mg\,{\sc ii}} } 
\newcommand{\mgi}{\mbox{Mg\,{\sc i}} } 
\newcommand{\mnii}{\mbox{Mn\,{\sc ii}} } 
\newcommand{\feii}{\mbox{Fe\,{\sc ii}} } 
\newcommand{\ciii}{\mbox{C\,{\sc iii]}} } 
\newcommand{\civ}{\mbox{C\,{\sc iv}} } 
\newcommand{\SiIII}{\mbox{Si\,{\sc iii}} } 
\newcommand{\AlIII}{\mbox{Al\,{\sc iii}} } 
\newcommand{\Lya}{\mbox{Ly\,{\sc $\alpha$}}} 
\newcommand{\zncr}{\mbox{Zn\,{\sc ii}\,$+$\,Cr\,{\sc ii}}} 
\newcommand{\znmg}{\mbox{Zn\,{\sc ii}\,$+$\,Mg\,{\sc i}}} 
\newcommand{\pks}{\mbox{PKS\,0405$-$385 }}
\newcommand{\pksx}{\mbox{PKS\,0405$-$385}}

\title{One sightline, many systems: a FLASH discovery of \hi towards scintillating quasar PKS 0405-385}

\author{E. F. Kerrison\orcidlink{0000-0002-0011-6922}}
\affiliation{Sydney Institute for Astronomy, School of Physics A28, University of Sydney, NSW 2006, Australia}
\alsoaffiliation{ATNF, CSIRO Space and Astronomy, PO Box 76, Epping, NSW 1710, Australia}
\alsoaffiliation{ARC Centre of Excellence for Gravitational Wave Discovery (OzGrav), Australia}
\email[E.F. Kerrison]{emily.kerrison@sydney.edu.au}

\author{H. Yoon\orcidlink{0000-0003-4048-2203}}
\affiliation{Institute for Data Innovation in Science, Seoul National University, 1 Gwanak-ro, Gwanak-gu, Seoul 08826, Republic of Korea}
\alsoaffiliation{Astronomy Program, Department of Physics and Astronomy, Seoul National University, 1 Gwanak-ro, Gwanak-gu, Seoul 08826, Republic of Korea}
\alsoaffiliation{Sydney Institute for Astronomy, School of Physics A28, University of Sydney, NSW 2006, Australia}
\alsoaffiliation{Korea Astronomy and Space Science Institute, 776 Daedeokdae-ro, Daejeon 34055, Republic of Korea}

\author{E. M. Sadler\orcidlink{0000-0002-1136-2555}}
\affiliation{Sydney Institute for Astronomy, School of Physics A28, University of Sydney, NSW 2006, Australia}
\alsoaffiliation{ATNF, CSIRO Space and Astronomy, PO Box 76, Epping, NSW 1710, Australia}
\alsoaffiliation{ARC Centre of Excellence for Gravitational Wave Discovery (OzGrav), Australia}

\author{Y. Kang\orcidlink{0000-0002-5261-5803}}
\affiliation{Kavli Institute for Particle Astrophysics and Cosmology, SLAC National Accelerator Laboratory, Stanford University, 2575 Sand Hill
Road, Menlo Park, CA 94025, USA}

\author{P. G. Edwards\orcidlink{0000-0002-8186-4753}}
\affiliation{ATNF, CSIRO Space and Astronomy, PO Box 76, Epping, NSW 1710, Australia}

\author{A. Tuntsov}
\affiliation{Manly Astrophysics, 15/41-42 East Esplanade, Manly, NSW 2095, Australia}

\author{J. P. Pritchard \orcidlink{0000-0003-1575-5249}}
\affiliation{ATNF, CSIRO Space and Astronomy, PO Box 76, Epping, NSW 1710, Australia}

\author{V. A. Moss \orcidlink{0000-0002-3005-9738}}
\affiliation{ATNF, CSIRO Space and Astronomy, PO Box 76, Epping, NSW 1710, Australia}
\alsoaffiliation{Sydney Institute for Astronomy, School of Physics A28, University of Sydney, NSW 2006, Australia}

\author{E. K. Mahony \orcidlink{0000-0002-5053-2828}}
\affiliation{ATNF, CSIRO Space and Astronomy, PO Box 76, Epping, NSW 1710, Australia}

\author{H. Bignall \orcidlink{0000-0001-6247-3071}}
\affiliation{Manly Astrophysics, 15/41-42 East Esplanade, Manly, NSW 2095, Australia}

\author{J.N.H.S. Aditya \orcidlink{0000-0002-0268-0375}}
\affiliation{Shanghai Astronomical Observatory, CAS, 80 Nandan Road, Shanghai 200030, P. R. China}
\alsoaffiliation{Sydney Institute for Astronomy, School of Physics A28, University of Sydney, NSW 2006, Australia}
\alsoaffiliation{ARC Centre of Excellence for All-Sky Astrophysics in 3 Dimensions (ASTRO 3D)}
\alsoaffiliation{State Key Laboratory of Radio Astronomy and Technology, A20 Datun Road, Chaoyang District, Beijing, P. R. China}

\author{J.R. Allison \orcidlink{0000-0003-0436-4680}}
\affiliation{First Light Fusion Ltd., Unit 9/10 Oxford Pioneer Park, Mead Road, Yarnton, Kidlington OX5 1QU, UK}

\author{S. Curran \orcidlink{0000-0002-1163-010X}}
\affiliation{Victoria University of Wellington School of Chemical and Physical Sciences, Ground floor Laby Building, Kelburn Parade, Wellington, NZ 6012}

\author{R. D. Ekers \orcidlink{0000-0002-3532-9928}}
\affiliation{ATNF, CSIRO Space and Astronomy, PO Box 76, Epping, NSW 1710, Australia}

\author{M. Glowacki \orcidlink{0000-0002-5067-8894}}
\affiliation{Institute for Astronomy, University of Edinburgh, Royal Observatory, Edinburgh, EH9 3HJ, United Kingdom}
\affiliation{Inter-University Institute for Data Intensive Astronomy, Department of Astronomy, University of Cape Town, Cape Town, South Africa}

\author{J. Stevens}
\affiliation{ATNF, CSIRO Space and Astronomy, PO Box 76, Epping, NSW 1710, Australia}

\author{R. Su}
\affiliation{Shanghai Astronomical Observatory, Chinese Academy of Sciences, 80 Nandan Road, Shanghai 200030, China}

\author{M. Whiting \orcidlink{0000-0003-1160-2077}}
\affiliation{ATNF, CSIRO Space and Astronomy, PO Box 76, Epping, NSW 1710, Australia}

\received {08 Dec 2025}
\revised  {04 Mar 2026}
\accepted {dd Mmm YYYY}
\published{xx Mmm 2026}

\begin{document}

\begin{abstract}
We report the discovery of an intervening 21\,cm absorption line at $z=0.882$ towards the $z=1.284$ quasar \pksx, identified in the First Large Absorption Survey in \hi (FLASH). This quasar once displayed the most rapid known intraday variability at radio frequencies, from which it earned the title of `the smallest radio quasar'. Although its size was revised upwards soon after based on updated scattering theory, \pks remains an important probe of Galactic plasma, and now also of intervening gas discovered through \hi absorption. We present new long-slit spectroscopy spanning both \pks and the candidate host of the intervening \hi gas. We identify \mgii and \feii absorption lines in this spectrum consistent with the redshift of the intervening \hix, as well as two additional, independent metal-line systems at $z= 0.907$ and $z=0.966$, but we cannot accurately pinpoint the host(s) of this intervening gas in current data. We revisit the radio variability of \pks in light of advances in scintillation theory, as well as extended monitoring with the Australia Telescope Compact Array and the Australian SKA Pathfinder, and find a revised linear size $\geq0.3\,$pc, but no new evidence of repeating intraday variability. 
\vspace{-5mm}
\end{abstract}

\section{INTRODUCTION}\label{sec:introduction}

Sightlines towards background AGN are important probes of the circumgalactic environment, providing a pencil beam sample of all the multi-phase gas along the line of sight. In particular, Lyman-$\alpha$ absorption lines in the restframe UV provide most of our current knowledge about neutral atomic hydrogen (\hix) in the distant Universe \citep[e.g.][]{wolfe05}. 

Radio measurements of 21\,cm absorption can also provide information about \hi in the distant Universe, particularly at $z<1.7$ where Lyman-$\alpha$ is not yet redshifted enough to be detectable by ground-based optical instruments, complicating the study of \hix-rich Damped Lyman-$\alpha$ (DLA) systems with $N_\text{HI} \geq 2\times10^{20}\,\text{cm}^{-2}$ \citep[e.g.][]{kanekar04,morganti15}. The optical depth of the 21\,cm absorption line is inversely related to the gas excitation (spin) temperature, so the \hi line is most sensitive to cold neutral gas with a spin temperature below $\sim300$\,K \citep{Morganti2018}.

In this paper, we report the discovery of redshifted 21\,cm \hi absorption along the line of sight to an intellectually (and physically) scintillating radio quasar, \pksx. The layout of the paper is as follows. In Section~\ref{sec:pks-history} we summarise what was known about \pks before the discovery of the intervening \hi presented here, including estimates of its angular size, and historical importance. In Section~\ref{sec:FLASH} we present the new \hi detection towards this source, and briefly discuss its line characteristics. Section~\ref{sec:new-science} offers a more complete analysis of the nature of \pks and the intervening gas based on new optical data (Section~\ref{sec:spectroscopy}) and an up-to-date radio lightcurve (Section~\ref{sec:radiolightcurve}). We discuss a few theoretical considerations on the propagation of the light from \pks through foreground matter in Section~\ref{sec:new-theory}. Throughout this work, we adopt a flat, $\Lambda$ cold dark matter ($\Lambda$CDM) cosmology in line with values from \citet{Planck2018}; $\Omega_{\rm{m}}$ = 0.315, $\Omega_\Lambda = 0.685$, and $H_0 =67.4\,\text{km\,s}^{-1}\text{Mpc}^{-1}$. 

\section{THE HISTORY OF \pks} \label{sec:pks-history}

\pks (J0406$-$3826) is an $m_i = 18.10\,$mag quasar at $z = 1.28$ primarily known today as an early and extreme example of an Intraday Variable (IDV) radio source. In observations taken with the Australia Telescope Compact Array (ATCA) in 1996, it showed hour-to-hour variations as high as $50\%$ at $4.8\,$GHz, an order of magnitude more extreme than the variability seen in the previously most variable IDV source, OJ 287 \citep{Kedziora-Chudczer1997}. At the time, scattering theory suggested such variations could only be produced towards a background source with an angular size less than 5$\,\mu$as, which would have made this the smallest known radio quasar \citep{Kedziora-Chudczer1997}. 
The initial IDV activity lasted several months, with a second period of IDV observed in 1998 \citep{Lucyna2001,Lucyna2006}. During the second period of IDV, 4 nights of optical observations were conducted by H.B. on the ANU 40\,inch Telescope at Siding Springs Observatory. The data were also reduced by H.B. using standard IRAF procedures \citep{Tody1993}. A standard deviation in relative photometry between \pks and a comparison star of similar magnitude at R-band was 0.08 magnitudes based on 11 measurements taken between 7 December - 10 December of that year. This variability was not deemed significant at the time and the data went unpublished, it will be discussed more thoroughly in a future work, along with other archival data on \pksx.
No further episodes of IDV were seen in a monitoring program that continued until 2002 April \citep{Lucyna2006}, and a few years later, \citet{Rickett2002} presented a revised distance to the Galactic scattering screen responsible for IDV, increasing the source size estimation.

 IDV notwithstanding, \pks is a strong ($S_{5\text{\,GHz}}\sim2\,$Jy) source used as part of the International Celestial Reference Frame \citep{Charlot2020}, and at high energies, it is a Fermi GeV gamma-ray blazar. \cite{Gong2022} noted quasi-periodic outbursts on a $\sim$2.8\,year timescale in Fermi data between 2008 August and 2021 November, however that trend has not continued in more recent years \citep{Abdollahi2023}. A visual examination of both long term radio monitoring with the Australia Telescope Compact Array, and v-band optical monitoring with the ASAS-SN network also shows several outbursts at both radio and optical wavelengths over this period \citep{Stevens2012, Kochanek2012}. These outbursts are physically unrelated to any IDV at radio frequencies; such multiwavelength flaring is often seen in blazars, and is typically explained by shocks forming in the core and propagating out along the radio jet \citep[e.g.][]{Beaklini2017}.

In the optical-IR, \pks is not red by the definition of \citet{Ross2015ExtremelySpectra} ($r_{AB} - W4_{Vega} > 14\,$mag), having $r_{AB} - W4_{Vega} = 11.3$\,mag, nor is it red using the more relaxed definition of \citet{Glowacki2019AnQuasars} (W2$-$W3 $> 3.5$), as it has a WISE colour W2$-$W3 $=2.56$\,mag. Thus although there is a high \hi detection rate towards red quasars, this is not one such source \citep{Carilli1998RedshiftedQuasars, Glowacki2019AnQuasars, Dutta2020UGMRTQuasars}. Also in the optical, this quasar has only one published spectrum, from \cite{Veron1990}, in which it is identified as a $z=1.285$ quasar based on \mgii and \ciii emission. An earlier spectrum was discussed by \cite{Savage1981}, who incorrectly placed the source at $z=2.04$ based on (mis)identifications of \civ and \Lya. However, this earlier spectrum was not published. 
\cite{Veron1990} note the presence of absorption lines in their spectrum, but state that the resolution is insufficient to attempt identification.
There is passing mention in \cite{Kedziora-Chudczer1997} of an intervening absorption system at $z=0.875$ in the \cite{Veron1990} spectrum, identified by R.W.\ Hunstead in 1996, presumably based on the association of absorption features in the \cite{Veron1990} spectrum with Fe\,{\sc ii} and Mg\,{\sc ii}. We present in the following section the first secure identification of this intervening system.

\section{A NEW DISCOVERY : INTERVENING \hi }\label{sec:FLASH}

Two intervening 21\,cm \hi lines were detected in a radio spectrum of \pks taken on 21 March, 2024 as part of the First Large Absorption Survey in \hi \citep[FLASH;][]{Allison2022,Yoon2025} conducted with the Australian SKA Pathfinder (ASKAP). FLASH is an untargeted search for \hi at redshifts $0.42 < z < 1$ towards all bright ($S\geq 30\,$mJy) radio sources in the southern sky excluding the Galactic plane. The FLASH spectral cubes have a $\sim30\,$arcsec spatial and 18.5\,kHz spectral resolution, and each FLASH spectrum uses the full 288\,MHz instantaneous bandwidth of the ASKAP radio telescope at 712--1000\,MHz \citep{Hotan2021}. A one dimensional spectrum is automatically extracted towards each source above the chosen flux density threshold, averaged over the beam, and continuum subtraction is performed in both the visibility and image plane as part of the ASKAPsoft pipeline \citep{Allison2022}. 

\begin{figure}
    \centering
    \includegraphics[width=0.99\linewidth]{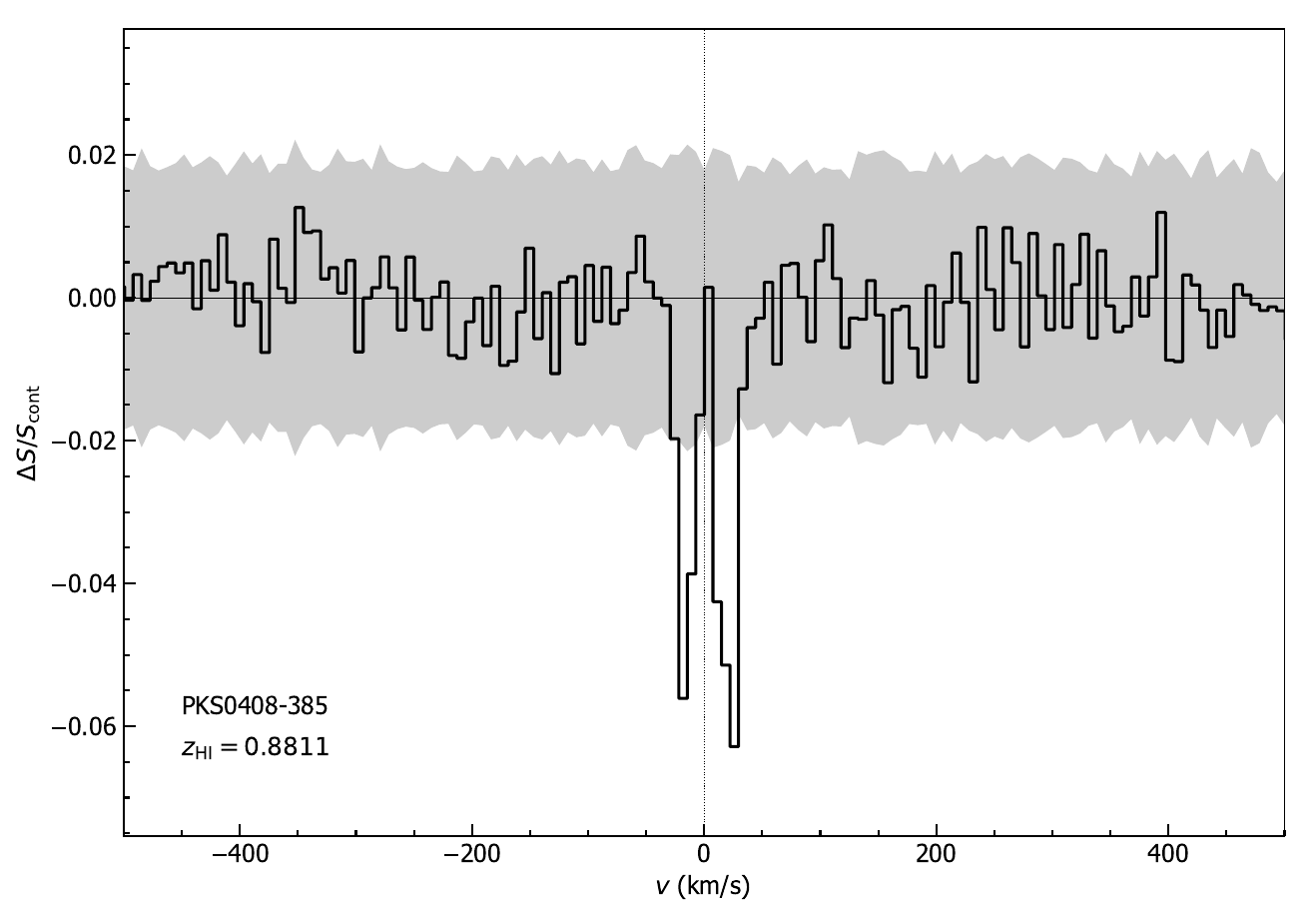}
    \caption{ASKAP spectrum of the intervening \hi lines towards \pks. The velocity scale is relative to the systemic redshift of $z=0.88115$. The y-axis indicates the absorption strength as a fraction of the continuum flux density. The grey band indicates $5\times$ the per-channel noise, taken from a blank sky spectrum around the target.}
    \label{fig:FLASH-spec}
\end{figure}

The spectrum of \pks is available from the public archive\footnote{The CSIRO ASKAP Science Data Archive (CASDA); https://research.csiro.au/casda/}, where it is listed as component 3a of scheduling block SB\,60306 (FLASH field 212). The segment of the spectrum containing the intervening \hi detection is shown in Figure~\ref{fig:FLASH-spec}, where two narrow components are clearly visible with a peak-to-peak velocity separation of approximately 45\,km\,s$^{-1}$. The characteristics of these lines are outlined in Table~\ref{tab:flashfinder-comps}, and were obtained using \texttt{FLASHfinder} \citep{Allison2012}.

\begin{table}[t]
\centering
\def\arraystretch{1.1}
\begin{tabular}{@{}lll@{}}
\hline
FLASH \hi parameters & Component 1 & Component 2 \\
\hline
Redshift $z$ & $0.8809853\pm0.0000004$  & $0.881207\pm0.000003$ \\
$\tau_\text{peak}$ & $0.059\pm0.007$ & 
$0.063\pm0.004$\\
$\tau_\text{int}$ (km\,s$^{-1}$) & $0.97\pm0.09$ & 
$1.29\pm0.07$ \\
$\Delta v$ (km\,s$^{-1}$) & $16.44\pm3.5$ & $20.4\pm1.1$ \\
ln (B) & 120 & 230 \\
N$_{\rm \hi}$ (T$_\text{s}$ = 100\,K) & $1.8\times 10^{20}$ cm$^{-2}$ & $2.4\times 10^{20}$ cm$^{-2}$ \\
N$_{\rm \hi}$ (T$_\text{s}$ = 1000\,K) & $1.8\times 10^{21}$ cm$^{-2}$ & $2.4\times 10^{21}$ cm$^{-2}$\\
\hline
\multicolumn{3}{c}{peak-to-peak separation: $44.1\,\text{km\,s}^{-1}$}\\
\hline
\end{tabular}
\caption{\hi linefinder measurements for \pksx, derived from fitting a simple Gaussian profile to each component. The first five rows correspond to output from the linefinder, the redshift ($z$) peak and integrated optical depths ($\tau_\text{peak}$, $\tau_\text{int}$), the velocity width ($\Delta v$) calculated as $\Delta v = \tau_{int}/\tau_{peak}$ and the logarithm of the Bayes factor, a statistical measure of the preference for a line existing at this location in the spectrum ($\ln\,(\text{B})$). The column density in the last two rows is derived using the familiar equation, $N_{\text{HI}} = 1.823 \times 10^{18} T_s \times f^{-1}\int\tau(\nu)d\nu$ and assuming covering factor $f=1$ and two different spin temperatures for the gas.} 
\label{tab:flashfinder-comps}
\end{table}

\begin{figure*}
    \centering
    \includegraphics[width=0.99\linewidth, trim = {1.8cm, 3.5cm, 1.8cm, 4cm}, clip]{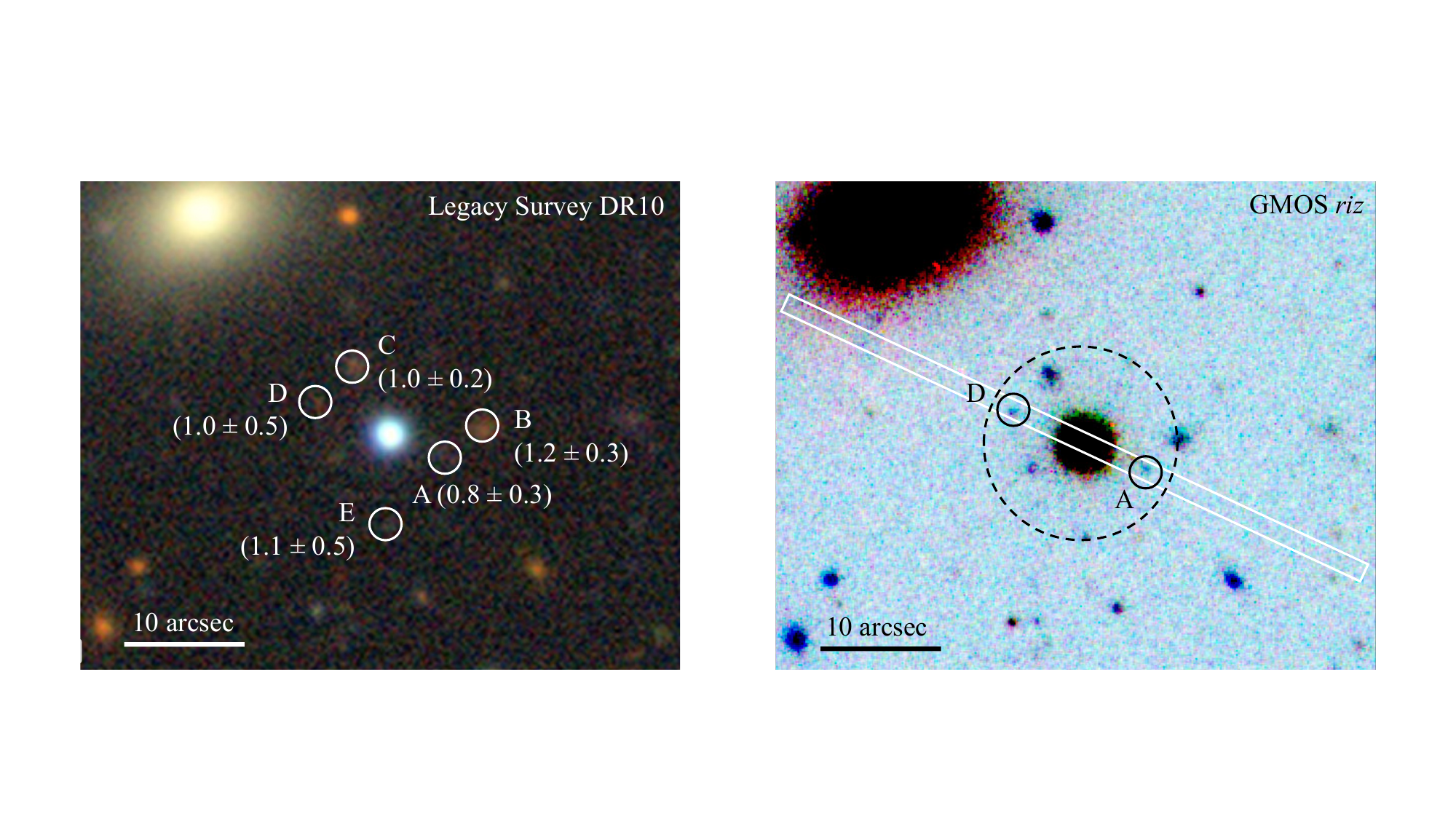}
    \caption{Left: three colour image taken from DR10 of the Legacy Survey \citep{Dey2019} of a region centred on \pksx. Five nearby galaxies visible in the image are identified as A--E. Galaxies A--D all have photometric redshifts from DR9 of the Legacy Survey within the range $\left[0.8, 1.2\right]$ as indicated in the image, with Galaxy A closest to the redshift of the FLASH detection at $z = 0.8\pm0.3$. Right: three colour image from Gemini GMOS obtained as part of follow up on this source. The white rectangle indicates the positioning of the slit used to obtain spectroscopy, aligned to span both \pks and Galaxy A coincidentally also spanning Galaxy D). The circle indicates a region of radius 50\,kpc at $z = 0.881$, the redshift of the FLASH detection, centred on \pksx. }
    \label{fig:img-compare}
\end{figure*}

The estimated \hi column densities suggest that this is a DLA system, and we note that its column only increases if we assume $\text{T}_\text{s}>100\,$K, as is likely at large galactic radii. This quasar-DLA pair is therefore a potential analogue to the intervening system detected in \hi towards PKS\,1127-2145 \citep{Kanekar2001}, where the \hi profile was seen to vary in optical depth over the course of 6 months, later attributed to scintillation caused by Galactic scattering \citep{Macquart2005}. It should also be measured against the quasar-DLA pair seen towards PKS\,2355-106. There, a second, intervening \hi absorption component separated from the first by $\sim55$\,km\,s$^{-1}$ appeared between initial GMRT observations in 2006, and follow up with both MeerKAT and GMRT in 2010. This variability in the \hi profile has been interpreted as the product of proper motion of the background source, since the quasar is canonically compact at VLBI resolution and showed insufficient variability in its radio continuum for scintillation to explain the observed variations in \hi optical depth \citep{Srianand2022EmergencePKS2355-106}. The lensed system PMN\,J0134-0931 is another useful comparison, shown in \citet{Kanekar2003} to exhibit a \hi absorption profile with two strong, narrow components separated by $\sim250\,\text{km\,s}^{-1}$ which is reproducible in models with a single, intervening galaxy disk sampled at discrete locations by separate, high surface brightness components of the background radio source. Moreover, recent high resolution, L-band follow up of twelve FLASH detections with the Very Long Baseline Array (VLBA) revealed eleven sources with complex or extended structure on milliarcsecond scales, including the only detection in that sample to have a two component profile, PKS\,2007$-$245 \citep{Aditya2025}. The velocity separation between the components there was $25\,\text{km\,s}^{-1}$. Although not an exact analogue (PKS\,2007$-$245 has lobes spanning $\sim40\,$mas), the velocity separation between the two \hi components here may similarly suggest an underlying complex or core-jet structure in the radio continuum source, where each component draws a discrete sightline through the intervening \hi gas.  

Most crucially though, the redshift of this intervening system ($z\approx0.881$) corresponds nicely with the redshift of the intervening line ($z\approx0.875$) which was identified some 30 years ago in the original \cite{Veron1990} spectrum, but never followed up.

\section{\pks REVISITED : NEW OBSERVATIONS}\label{sec:new-science}

Spurred on by this new detection of intervening \hi from an untargeted search, we have revisited \pks to see what can be learnt about both the quasar and this intervening system with additional radio and optical data.

Since the redshift of the newly-discovered \hi gas aligns closely with that of the absorption line reported in the original optical spectrum, it offers potential new insight into the multi-phase interstellar (or circumgalactic) medium of the intervening system, if this system can be more securely identified.

Images from DR10 of the Legacy Survey  \citep{Dey2019} reveal five nearby galaxies labelled counter-clockwise A--E in Figure~\ref{fig:img-compare}, left, which have DR9 photometric redshifts broadly commensurate with the FLASH detection given their uncertainties $( 0.8 < z_{\text{phot}} < 1.2)$. Out of these candidates, Galaxy A has the closest redshift to the FLASH detection at $z = 0.8 \pm 0.3$, making it the most likely host of the \hi gas. Furthermore, Galaxy A has the bluest optical colours of these candidates ($g - i = 0.58$\,mag from the Legacy Survey DR10) and is therefore likely starforming, so a high \hi mass would not be surprising. Should this indeed be the host, the quasar sightline is passing through gas at an impact parameter of $\sim36\,$kpc ($5.07\,$arcsec), probing the circumgalactic medium at a distance where strong absorption lines are common, at least at earlier cosmic times \citep{Adelberger2005}. Alternatively, the \hi may exist in the ISM of another galaxy for which the light in the legacy images is entirely blended with that of the background quasar, due to its extremely small impact parameter. As a third and final alternative, galaxies A--E may form a foreground group, in which case the \hi detected in FLASH may sample a clumpy, extragalactic medium, evidence of galaxy-galaxy interactions \citep{Weng2022}. Unfortunately, the Legacy Survey photometric redshifts are not accurate enough to distinguish between these pictures; optical spectroscopy is required. To achieve this, we obtained fast turnaround time on the 8.1\,m Gemini South telescope, using the Gemini Multi-Object Spectrograph (GMOS) for both optical imaging and spectroscopy under project GS-2024B-FT-215 (P.I. Yoon). An analysis of this new optical data is presented below.

\subsection{GMOS imaging}\label{sec:imaging}
Our imaging observations comprise $3\times100$\,s in each of the \textit{r} and \textit{i} bands, and $3\times80$\,s in the \textit{z}-band, all obtained on December 10th, 2024. Data pre-processing and reduction were performed using DRAGONS \citep[Data Reduction for Astronomy from Gemini Observatory North and South;][]{Labrie2023}. We re-identified galaxies A--E in the resulting three colour composite image, shown in Figure~\ref{fig:img-compare}, right. No additional nearby galaxies were identified in the GMOS images. Also shown in Figure~\ref{fig:img-compare}, right is the alignment used for longslit spectroscopy (253.0 deg E of N), to ensure both \pks and the most likely candidate, Galaxy A, would be captured. The circles centred on \pks represent a region of radius 50\,kpc at $z = 0.881$, the redshift of the FLASH detection.

\subsection{GMOS spectroscopy}\label{sec:spectroscopy}

\begin{figure*}
    \centering
    \includegraphics[width=0.9\linewidth]{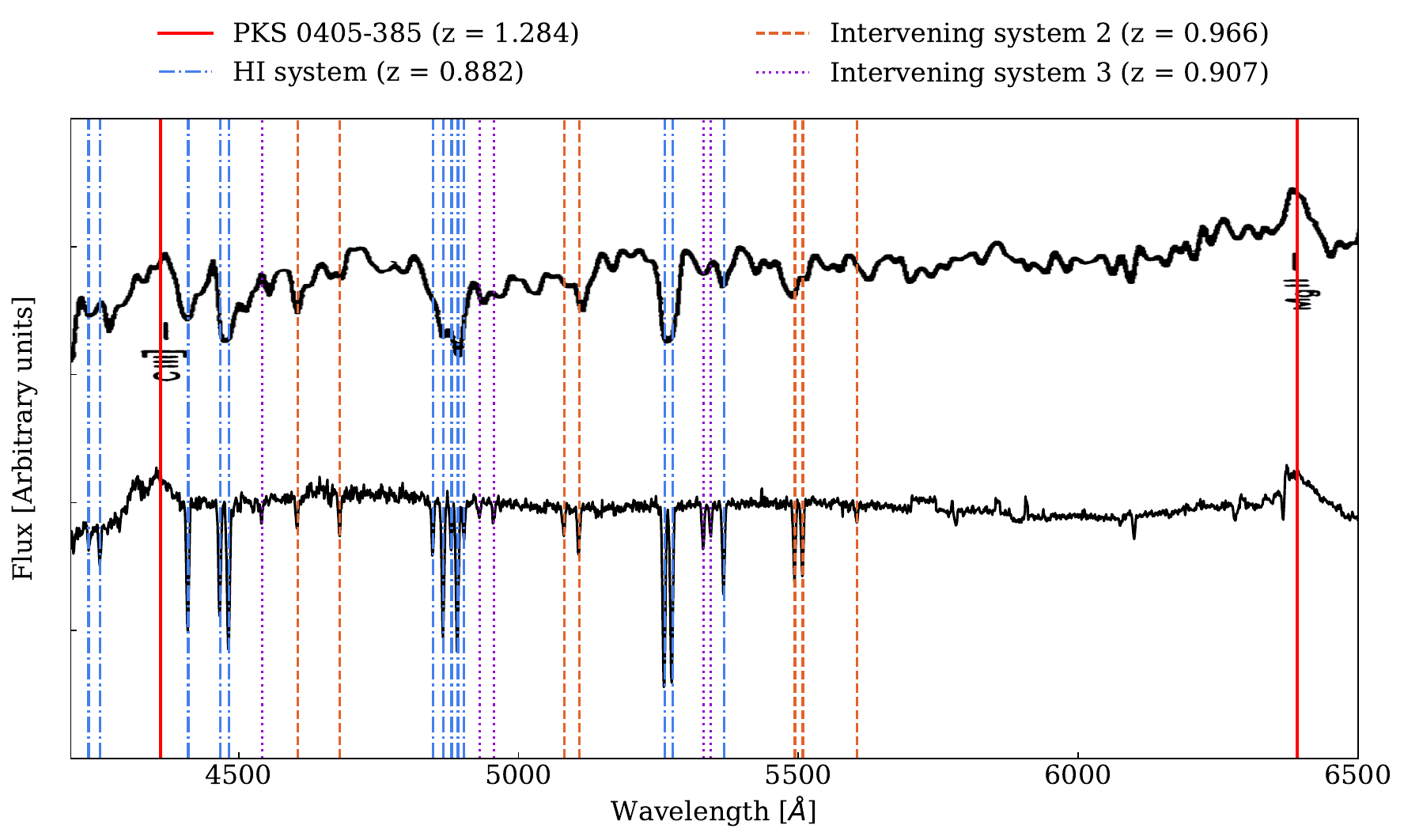}
    \caption{The original optical spectrum from \citep{Veron1990} (top) compared to our new spectrum taken with GMOS-S (bottom). Vertical lines indicate emission lines associated with background quasar \pks (red, solid), absorption lines associated with the intervening galaxy detected in FLASH data (blue, dot-dashed), and two further, previously unidentified intervening galaxies (orange, dashed and violet, dotted). Lines were identified using MARZ and the new Gemini spectrum only. Nevertheless, a number of lines from both intervening systems are visible in the original \cite{Veron1990} spectrum.}
    \label{fig:spec-compare}
\end{figure*}

Spectroscopic observations were carried out on December 9th, 2024 under arcsecond seeing conditions using the B600 grating ($R\sim 1688$, or $\sim0.8$\,nm FWHM at 510\,nm) and a slit width of 1.5\,arcseconds. Data acquisition was split into $4 \times 900$\,s exposures, binned $2 \times2$ in both directions, with two sets using a central wavelength of 510\,nm and two set to 520\,nm, to fill the gap between detectors. The spectra have a resulting wavelength coverage of approximately $3550-6780$\,\AA.

\begin{table*}[h!]
\centering
\def\arraystretch{1.}
\begin{tabular}{@{}llllllllllllll@{}}
\hline
& & & \multicolumn{2}{l}{\pks} & \multicolumn{2}{l}{\hi system} & \multicolumn{2}{l}{System \#2}   & \multicolumn{2}{l}{System \#3} \\
& & & \multicolumn{2}{l}{$z = 1.284$} & \multicolumn{2}{l}{$z = 0.882$} & \multicolumn{2}{l}{$z = 0.966$}  & \multicolumn{2}{l}{$z = 0.907$} \\
\hline
Type & Transition & $\lambda_0$ & $\lambda_{\text{obs}}$  &  $EW_{\text{rest}}$  &  $\lambda_{\text{obs}}$  &  $EW_{\text{rest}}$ & $\lambda_{\text{obs}}$  &  $EW_{\text{rest}}$ & $\lambda_{\text{obs}}$  &  $EW_{\text{rest}}$ \\
 & & ({\AA}) & ({\AA}) & ({\AA}) & ({\AA}) & ({\AA}) & ({\AA}) & ({\AA}) &({\AA}) & ({\AA}) \\
\hline
\hline
Emission & \mgii & 2803.53 &  &  &  \\
& \mgii & 2796.35 & \multirow{-2}{*}{6396.2} & \multirow{-2}{*}{$8.6\pm0.4$} & \\
& \ciii & 1908.73 & 4353.4 & $4.0\pm0.9$  \\
\hline
Absorption & \mgi & 2852.13 & & & 5367.5 & $0.93\pm0.07$ & 5605.1 & $0.25\pm0.08$ & \\
& \mgii & 2803.53 & & & 5261.1 & $2.12\pm0.07$ & 5508.0 & $0.69\pm0.07$ & 5344.0 & $0.42\pm0.07$ \\
& \mgii & 2796.35 & & & 5275.1 & $2.25\pm0.07$ & 5494.0 & $0.73\pm0.07$ & 5330.6 & $0.28\pm0.07$\\
& \mnii & 2606.51 & & & 4903.0 & $0.40\pm0.12$ \\
& \feii & 2599.40 & & & 4891.9 & $1.89\pm0.12$ & 5109.0 & $0.52\pm0.08$ & 4956.3 & $0.20\pm0.11$\\
& \mnii & 2593.72 & & & 4880.6 & $0.51\pm0.12$ \\
& \feii & 2585.87 & & & 4865.6 & $1.53\pm0.12$ & 5081.6 & $0.30\pm0.08$ & 4930.5 & $0.24\pm0.11$\\
& \mnii & 2576.10 & & & 4846.8 & $0.58\pm0.12$  \\
& \feii & 2382.04 & & & 4482.4 & $1.84\pm0.14$ & 4680.6 & $0.38\pm0.12$ & \\
& \feii & 2373.74 & & & 4466.7 & $1.22\pm0.14$ \\
& \feii & 2344.28 & & & 4410.0 & $1.37\pm0.15$ & 4606.0 & $0.28\pm0.12$ & 4541.7 & $0.16\pm0.13$ \\
& \feii & 2260.78 & & & 4252.0 & $0.42\pm0.11$\\
& \feii & 2249.87 & & & 4232.0 & $0.21\pm0.11$\\
\hline
\end{tabular}
\caption{Lines identified in the GMOS spectrum assigned to each system. We note that the \mgii doublet seen in emission at the redshift of \pks is not resolved. All $\lambda_\text{obs}$ values have a measurement uncertainty of $\pm0.05\,$\AA, and the redshifts should likewise be considered to have a measurement uncertainty of $\pm0.0005$.} 
\label{tab:optical-lines-wide}
\vspace{1cm}
\end{table*}

Data pre-processing and reduction were once again performed with DRAGONS, and RV correction to the heliocentric frame was performed using Astropy's \texttt{SpecCoord} class. Initial line and redshift identification was made with a modified version of MARZ \citep[Manual and Automatic Redshifting Software;][]{Hinton2016} with additional, restframe-UV lines. Equivalent Width (EW) measurements were performed with the \texttt{Specutils} package within Astropy \citep{Specutils}. The co-added, reduced spectrum has a median signal to noise ratio  $SNR=107$ per pixel across the full spectrum.

The new GMOS spectrum is shown in Figure~\ref{fig:spec-compare} alongside the original spectrum from \cite{Veron1990}. The full list of lines identified in the new spectrum is given in Table~\ref{tab:optical-lines-wide}. The signal-to-noise ratio was not sufficient to extract useable spectra towards the intervening system(s). In both spectra, the \mgii doublet and \ciii from \pks are visible in emission, indicated by the red solid lines in Figure~\ref{fig:img-compare}. Given the width of the \ciii feature, it is likely in fact to be a blend of \ciii ($\lambda$1908.83), \SiIII ($\lambda$1892) and \AlIII ($\lambda$1860), the ratios of which can provide insight into the Eddington accretion rate of the quasar \citep{Marziani2014, Martinez2018}. However, we leave further, detailed discussion of the emission line properties of \pks to future work, ideally with a higher resolution spectrum in which these blended features can be better resolved and subsequently modelled. 

\begin{figure*}[t]
    \centering
    \includegraphics[width=0.99\linewidth]{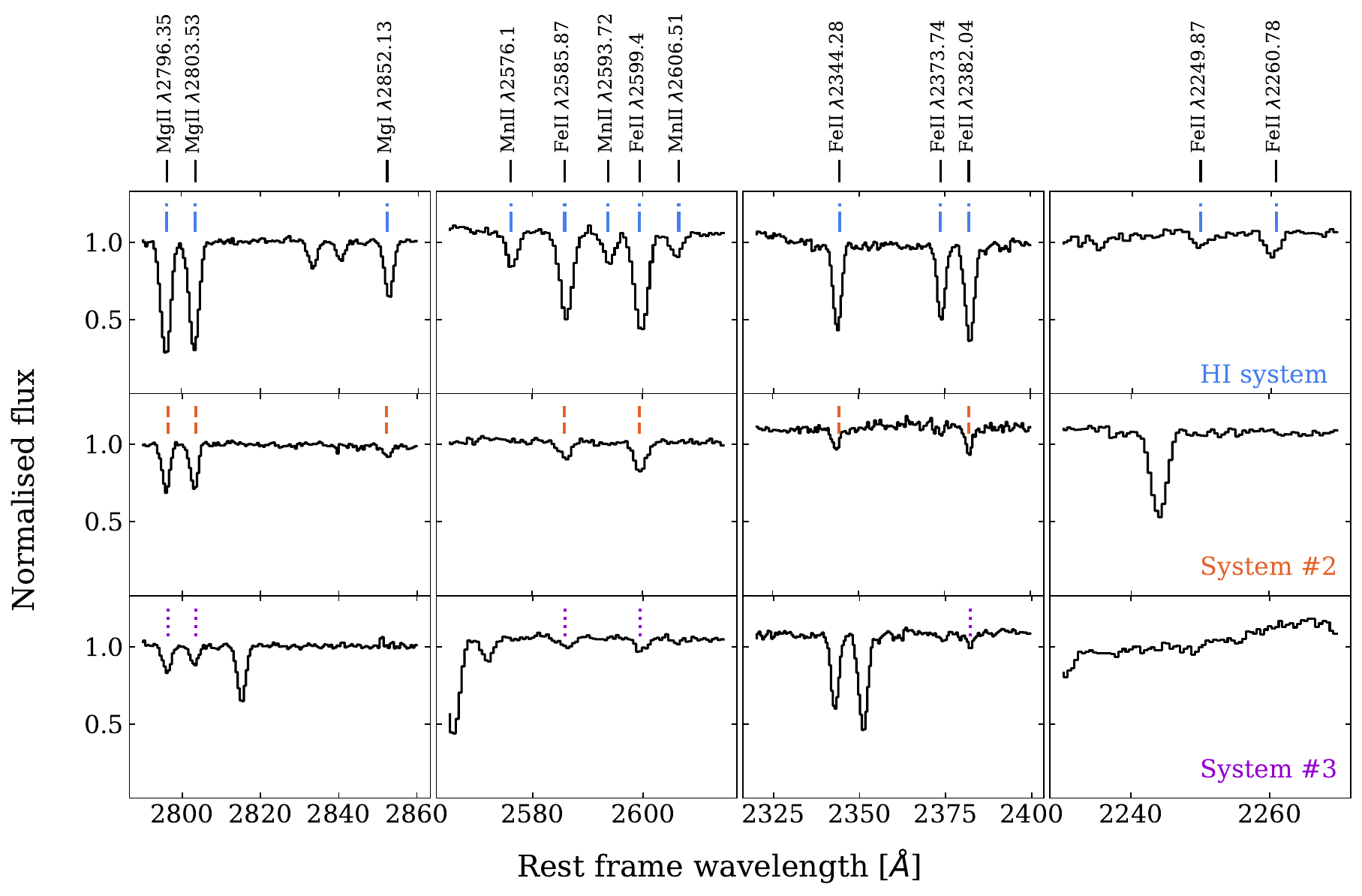}
    \caption{Cutouts from the continuum-subtracted GMOS spectrum presented in Figure~\ref{fig:spec-compare} centred on the regions in which absorption lines are seen at the redshift of the \hi system (top row) the second, intervening system at $z = 0.966$ (middle row) and the third at $z = 0.907$ (bottom row). Vertical lines in each subplot indicate the detection of an absorption line corresponding to the labels at the top of the figure.}
    \label{fig:spec-zoom}
\end{figure*}

The GMOS spectrum clearly resolves the \mgii doublet in absorption at $z = 0.882$ (violet dashed lines), which is unresolved but visible in the \cite{Veron1990} spectrum also. Furthermore, the GMOS spectrum shows several other absorption lines at this redshift, including iron lines at $\sim4400$\,\AA, and iron and manganese lines at $\sim4880$\,\AA, both of which also align with unresolved features in the \cite{Veron1990} spectrum. These suggest our original \hi detection is towards at least one galaxy at $z=0.882$ which we refer to henceforth as the `\hi system'. There are possibly also several more lines at the redshift of the FLASH \hi line, including \zncr ($\lambda$2062.66), and \znmg ($\lambda$2026.14). However, these fall at an observed wavelength $\lambda_o < 4300$\AA, where the noise in our new GMOS spectrum is higher, so these detections are currently of low significance. 

\newpage
Interestingly, we also identify two further intervening systems at $z = 0.966$ (`Intervening System 2', orange dotted-dashed lines) and $z=0.907$ (`Intervening System 3', violet dotted lines), for which several absorption features agree with low-significance features in the \cite{Veron1990} spectrum. Again, the GMOS spectrum shows the \mgii doublet, \mgi and several iron lines at the redshift of the second intervening system, and \mgii and \feii lines at the redshift of the third. Higher resolution cutouts around the detected lines for all three systems are provided in Figure~\ref{fig:spec-zoom}. The wavelength range of our spectrum ($3550-6780$\,\AA) does not extend to cover typical strong, nebular emission lines such as [OII], [OIII] or H$\beta$ at the redshift of any of our intervening systems, so we cannot search for these  either against the bright quasar spectrum, or at the location along the slit corresponding to Galaxy A. 

\vspace{-1mm}
A re-examination of the FLASH spectrum within $\pm500\,\text{km\,s}^{-1}$ of $z = 0.907$ and $z = 0.966$ reveals no \hi detection at the $3\,\sigma$ level. However, combining the root mean square noise in optical depth of the FLASH spectrum locally ($\tau_{\text{rms}} = 0.004$) with the average full width zero intensity of intervening \hi from \cite{Curran2021} ($108\,\text{km\,s}^{-1}$), we can place an upper limit on the amount of cold, neutral gas at the redshift of these second and third intervening systems. We estimate that they must each have $N_{\text{HI}} \leq 1.3\times10^{20}\,\text{cm}^{-2}$ ($\text{T}_\text{s} = 100\,$K, $f=1$), which would make these sub-DLA systems. We note here that the \hi system would be considered iron rich using the classification scheme of \citet{Dutta2017} ($EW_\text{rest} > 1.0\,$\AA), who found that such systems were four times more likely to exhibit \hi absorption than their iron poor counterparts (like intervening systems 2 and 3) at $0.5 < z < 1.5$.  However, to say more on the abundances of metals in this intervening gas requires higher spectral resolution and, for system 3 at least, higher signal to noise also.

Ultimately, our question as to the origin of the detected \hi remains unanswered. Since we were unable to extract any identifiable spectral features at the location of Galaxy A and \pksx, we were unable to either confirm or disprove this as the host of the intervening \hix. Furthermore, many absorption features in the GMOS spectrum are likely saturated making it impossible to deduce abundances, though such analysis would theoretically be possible with the combined detection of (unsaturated) metal lines and neutral \hi with well-constrained velocity dispersion.  Integral field spectroscopy spanning \pks and galaxies A--E, along with deeper optical imaging, will be key to securely identifying both intervening systems seen in absorption against \pksx, and further analysing their metallicity.

\subsection{Radio monitoring}\label{sec:radiolightcurve}

\begin{figure*}[h!]
    \centering
    \includegraphics[width=0.99\linewidth]{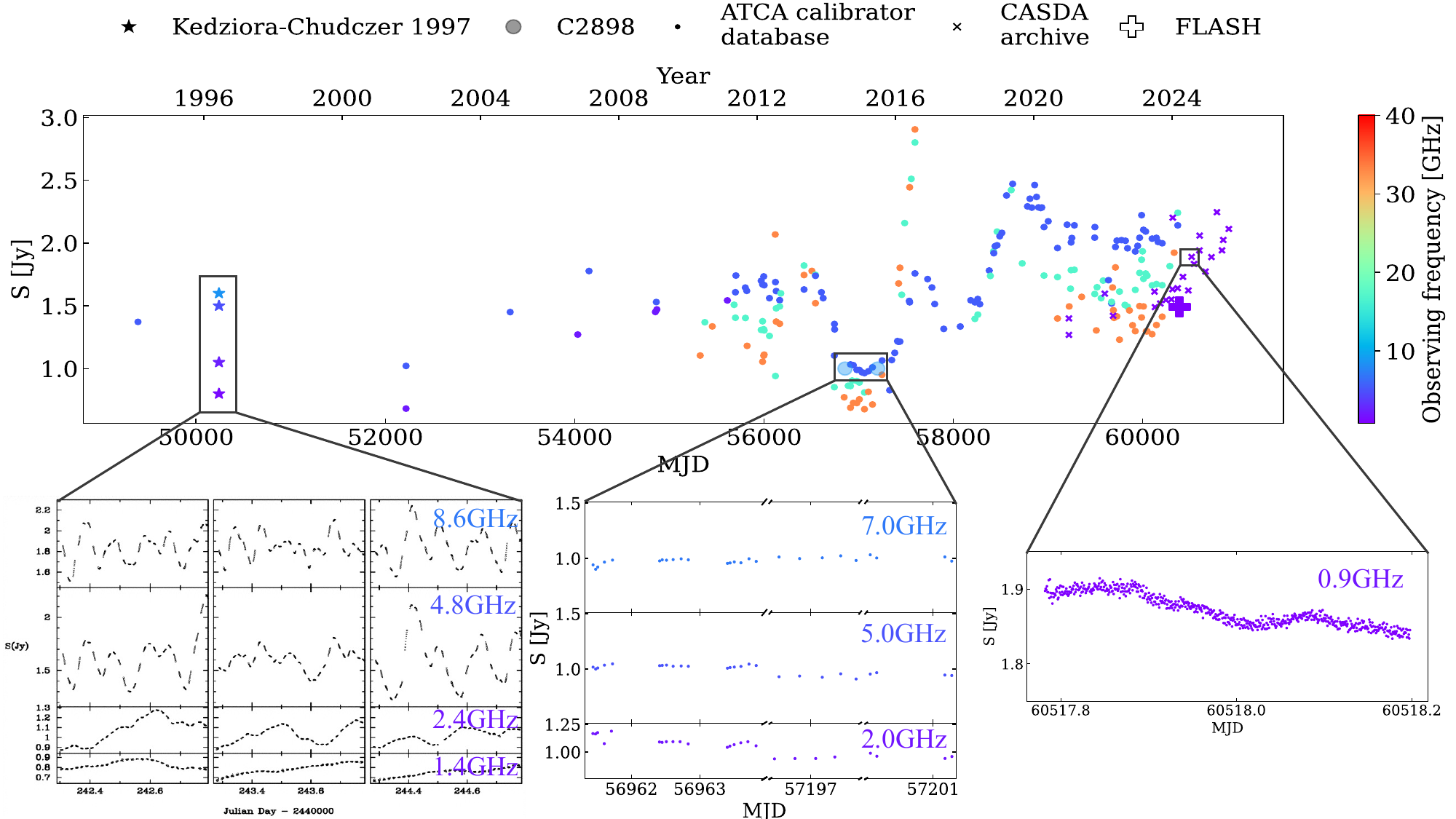}
    \caption{The radio lightcurve of \pks compiled from targeted monitoring programmes C007, and C1730 (filled circles) with the ATCA, labelled as `ATCA calibrator database'. We additionally show the original, broadband fluxes from \citet{Kedziora-Chudczer1997} (stars), with an inset showing the IDV detected during those observations (bottom, left), as well as a later ATCA monitoring programme C2898 during which IDV was not observed (larger, semi-transparent circles, middle inset). Further, coincidental observations of \pks are taken from the CASDA archive (crosses), including the FLASH observations (filled vertical cross), and a 10-hour pointing observed as part of the Evolutionary Map of the Universe survey \citep[EMU][]{Norris2011}.}
    \label{fig:radio-lc}
\end{figure*}

As discussed in Section~\ref{sec:introduction}, \pks is an interesting source itself, exhibiting both powerful, episodic IDV and intermittent $\gamma$-ray flares. As a result, it has been the subject of long-running radio monitoring, both targeted and incidental, which we compile and present here for the first time in Figure~\ref{fig:radio-lc} alongside the original IDV observations (lower-left inset). \pks has been regularly monitored with the Australia Telescope Compact Array over the frequency range 5 to 40\,GHz under observing programmes C007 and C1730 \citep{Stevens2012} since 2010 (filled circles in Figure~\ref{fig:radio-lc}). \pks was also one of the sources included in a search for intra-day variability at 2, 5, and 7\,GHz under observing program C2898 between July 2014 and June 2015 (larger, semi-transparent circles). No evidence for IDV was seen (lower-centre inset plot), and the longer term monitoring indicates the source was at its most quiescent over that year. However, just as that program ended, \pks underwent a rapid brightening, reaching historical high flux densities in mid-2016 in the 15\,mm and 7\,mm bands. 
\pks is also a calibrator source for the Atacama Large Millimetre Array (ALMA), and 90--240\,GHz observations show a peak in flux density coincident with this 2016 flare.\footnote{https://almascience.eso.org/alma-data/calibrator-catalogue}
We also re-imaged a 10-hour observation at 0.95\,GHz from the Evolutionary Map of the Universe survey \citep[EMU][]{Norris2011} taken with ASKAP in June 2025 using \texttt{dstools} \citep{dstools}. This is shown in the bottom right inset plot, where there is only minor variability at the level of 1 per cent over the course of several hours. However, we note that even in the original \citet{Kedziora-Chudczer1997} data, the variability was weakest below $2\,$GHz, so this does not place a strong constraint on the recent level of IDV in this source; higher frequency observations are needed.  
\cite{Lucyna2001} postulate that the IDV observed in 1996 ceased as a result of the increasing size of the scintillating component, although an alternative explanation could invoke changes in the properties of the scattering screen, as seen towards {PKS\,1257-326} \citep{Koay2011}. 

It is well established that radio flares in blazars are often accompanied by the ejection of a new parsec-scale jet component which can dominate the total flux density, and which can initially be sufficiently compact to produce IDV if a suitable scattering screen is present along the line of sight. Unfortunately, there was no IDV monitoring program in operation during the 2016 outburst to test this hypothesis, so broadband monitoring during and immediately after future flares would prove extremely useful in this regard.

At higher resolutions, \cite{Lucyna2001} derive an upper limit of 0.15\,mas on the size of the core based on 8.4\,GHz VLBA observations, which places an upper limit on the linear size of the core of 1.3\,pc. 
Jet components are also visible in their 2.3 and 8.4\,GHz images, extending up to 20\,mas from the core, but the limited number of observations preclude a reliable estimate of apparent jet component speeds.

The linear polarisation of \pks was also studied from the period of IDV observed in 1996, and \citet{Rickett2002b} determined the variations could be best explained by three, compact components forming an oblique source of $14\times20\,\mu\text{as}$ at 4.8\,GHz, corresponding to a linear size of approximately $0.3\,$pc at the source redshift. Current best estimates then put the \pks core at $0.3-1.3\,$pc. At the redshift of the \hix, the core emission has a linear extent of only $\sim1.5\,$pc, easily subtended by a typical \hi cloud which is thought to span $\sim$a few parsecs in local ($z\lesssim 0.1$), extragalactic systems \citep{Srianand2013Parsec-scale0.079,Gupta2018Revealing0.017}, or perhaps as much as $\sim10$\,pc in the analogous $z\sim0.3$ quasar-DLA pair PKS\,1127-2145 already discussed in Section~\ref{sec:FLASH} and presented in \citet{Kanekar2001}. Contemporary VLBI observations, ideally at or close to the frequency of the \hi detection as in \citet{Aditya2025}, will be crucial to understanding the structure of \pks, and may also provide insight into why it illuminates two \hi structures with discrete velocities. 


\section{\pks REVISITED : PROPAGATION EFFECTS}\label{sec:new-theory}

In light of our new observations presented in Section~\ref{sec:new-science}, we consider here a few possibilities as to how the intervening matter from the three systems might affect the propagation of light from \pksx.

\subsection{Could \pks be gravitationally lensed?}
We now know there are three intervening galaxies along the line of sight to \pks close enough in angular separation to produce absorption lines in its spectrum. We might therefore consider whether the mass along this sightline is sufficient to gravitationally lens the background emissions of the blazar. For a source at $z_s\approx1.284$ and lens $z\approx0.882$, the critical column density of matter required for strong gravitational lensing -- i.e., multiple imaging and/or significant magnification -- to occur is $0.89\,\mathrm{g}\,\mathrm{cm}^{-2}$, with our assumed cosmology. 

The surface density of neutral hydrogen seen in the \hi system is only $1\times10^{-3}\,\mathrm{g}\,\mathrm{cm}^{-2}$, almost three orders of magnitude below that required for lensing. Since we cannot constrain the abundances of the gas in the intervening \hi system from our current data, we cannot determine the complete gas surface density, let alone the total matter density along the line of sight. As a first approximation then, we can consider the total matter surface density in our own solar neighbourhood. \citet{McKee2015} put the local Galactic \hi density at $\sim2\times10^{-3}\,\text{g}\,\text{cm}^{-2}$, remarkably close to the \hi surface density detected in FLASH. They put the total matter surface density at $\sim1\times10^{-2}\,\text{g}\,\text{cm}^{-2}$, still two orders of magnitude below the gravitational lensing threshold if this were the density intersected at $z=0.882$. Of course the matter distribution in the \hi system might be entirely unlike our own Milky Way, and is almost certainly sampled by the \pks sightline at a different galactic radius, but in the absence of additional data it is impossible to say more. Furthermore, physical association between the three systems seen in absorption is unlikely, as they are separated along the line of sight by tens of Mpc (interpreting observed redshifts as cosmological) and in velocity by tens of thousands of $\text{km\,s}^{-1}$ (assuming instead that redshift differences are due to peculiar motion). Therefore, there is currently no evidence for any additional mass contribution at group or cluster scales. In short it is unlikely that the light from \pks is lensed by the foreground systems, but a better understanding of their baryonic matter content will help to better constrain this problem.

\subsection{What effect does Galactic scattering have?}\label{sec:galactic-scattering}
\citet{Macquart2005} showed that a multi-component \hi profile could appear variable due to Galactic scattering and propagation effects. We cannot say anything of the variability in our \hi system from one \hi observation, though followup during an episode of IDV would be particularly interesting to search for spectral line variability. Nevertheless, it is worth considering whether scintillation can offer further insights into the structure of \pksx, and the intervening gas.

In the original \citet{Kedziora-Chudczer1997} paper, the angular size of the \pks core was constrained to $< 5\,\mu$as, the Fresnel scale at which scintillation becomes significant for a scattering screen of Galactic plasma at a distance of 500\,pc. However a screen at 30\,pc is perfectly reasonable, and would require an angular diameter less than $20\,\mu$as; indeed screens at or below 10\,pc have since been observed \citep{2025NatAs...9.1053R, 2021MNRAS.502.3294W}, further relaxing the source size constraints to $38\,\mu$as, corresponding to a linear size of $\sim0.3$\,pc at $z=1.284$, similar to the size derived from VLBI in Section~\ref{sec:radiolightcurve}.

Of course, we now know from Section~\ref{sec:spectroscopy} that there are at least three intervening systems along the line of sight towards \pks, each with their own ISM. Therefore Galactic plasma is not the only possible source of scattering (or angular broadening); we must also consider how the medium of the three intervening systems might contribute to the angular source size and variability. 

\subsection{Could there be scattering from intervening systems?}\label{sec:scintillation-theory}
The theoretical breakthrough made by \cite{Macquart2005} as mentioned in Section~\ref{sec:galactic-scattering} was motivated by \citet{Kanekar2001}, who originally considered whether Interstellar Scintillation (ISS) caused by the ISM of an \textit{intervening} galaxy might cause intraday fluctuations in both the background radio continuum, and intervening \hi line profiles of a similar quasar-galaxy pair. We consider again whether the plasma in such intervening systems could contribute meaningfully to angular scattering, which would in turn minimise any observed ISS, and artificially increase the angular diameter of \pksx. 

\hi absorption traces a different phase of the ISM to that responsible for angular scattering (cold, neutral as opposed to ionised), but we can still use it to make a first order approximation on the expected level of scattering, provided that we assume some relationship between the two phases. We can derive one such relationship by looking at the pulse broadening of pulsars as a function of Galactic \hi column density. Using the scattering time measurements from the Australia Telescope National Facility (ATNF) pulsar catalogue \citep{2005AJ....129.1993M} and a model of Galactic \hi from \citet{Kalberla2009}, a \hi column of $\sim4.2\times10^{20}\,\text{cm}^{-2}$ as seen towards \pks might produce a scattering angle $\alpha \sim 0.1-1\,\text{mas}$ at 1\,GHz in our Galaxy, or something a factor of approximately 4 lower at $z = 0.882$, where it would correspond to intrinsically higher frequency and hence weaker scattering. At the frequencies at which IDV was observed and at the redshift of the FLASH detection, this drops to $\alpha\sim1-10\,\mu\text{as}$, just below the threshold required to quench the IDV produced by the Galactic screen discussed in Section~\ref{sec:galactic-scattering}, or significantly affect the angular size of \pksx. Since the \hi column density towards intervening systems 2 and 3 is even lower than the \hi system, this framework would suggest they contribute an even smaller scattering angle to the light coming from \pksx. We must reiterate that the above is only a first order approximation of the effect of these intervening screens. 

Once more, higher resolution optical spectroscopy would provide crucial insights into the mechanics of intergalactic scattering by allowing us to better constrain the multiphase gas along the line of sight. Coupled with further radio monitoring to detect new occurrences of IDV -- potentially with a time-domain study of the \hi profile -- this extra data may offer new insights into plasma physics from cosmological distances to our own, Galactic neighbourhood.

\section{SUMMARY}
\label{sec:conclusion}

A reconsideration of \pks shows it is in possession of a compact component $0.3-1.3\,$pc based on both VLBI imaging and a better understanding of Galactic scattering. The linear scales probed by the ISS intra-day variability are 0.3\,pc or less at the \hi  absorption frequency, so the IDV is likely to change across the HI absorption profile. Furthermore, long-term radio monitoring reveals several periods of rapid brightening indicative of episodic blazar activity from a compact core with structure seen on a scale of 1.3\,pc with VLBI at 8.4\,GHz. So, the HI absorption and its variability could be useful for ongoing studies of jet lifecycles.

Coincidentally, in an untargeted search for \hi absorption conducted as part of the ASKAP-FLASH survey, intervening \hi was detected towards \pks at $z=0.882$, corresponding to the redshift of absorption lines reportedly identified in the original optical spectrum of the source. Previous VLBI images and flares in ATCA monitoring suggest that the structure of the parsec-scale jet in \pks may have multiple components; a core and a bright jet component at the epoch of our FLASH observation could explain the two \hi features separated by $\sim45\,\text{km\,s}^{–1}$ if each continuum component samples a different region of an intervening disk, or an extragalactic, clumpy medium. However, more recent VLBI observations would be required to confirm the presence of such structure today. We obtained Gemini GMOS spectroscopy towards \pks and the potential host of the \hi gas to confirm the presence of the intervening system, and we identified the (likely) original, optical absorption lines at this redshift as the \mgii doublet, with additional \feii and \mnii absorption features revealed in the new spectrum. However, we could not confirm Galaxy A as the host of the \hi due to a lack of spectral resolution and signal-to-noise. Nevertheless, a number of other metal lines are also identifiable in the spectrum at this redshift, and we further identify the presence of two further, iron-rich intervening systems at $z = 0.907$ and $z=0.966$, which are not currently detected in \hix. The gaseous systems detected in intervening absorption are not likely to contribute to either IDV or scatter broadening of the background quasar. Scattering, and even interstellar scintillation in intervening galaxies \textit{does} have a noticeable effect on Fast Radio Bursts, which are extragalactic sources with an extremely small diameter. Comparison to \hi absorption in cases such as this one will be interesting, but is beyond the scope of this paper.

The evolution of metallicity at the redshifts probed by the intervening systems towards \pks is not well studied, largely due to a lack of optically-selected DLAs at these distances. This leaves the period just after cosmic noon critically under-explored, although we know star formation rates begin to decline here and gas distributions must therefore change \citep{Madau2014}. The case of \pks clearly demonstrates that radio selection of DLAs via intervening 21-cm absorption is a viable pathway to understanding metallicity evolution in this period. This technique will only grow in power with the progress of large-area, untargeted searches for \hi in absorption such as FLASH. In all such cases though, the most interesting science can only be extracted from these systems with sufficient multiwavelength data.  Repeat radio spectral observations will allow us to search for variability in the \hi absorption features which has seldom been detected \citep{Kanekar2001, Srianand2022EmergencePKS2355-106, allison2017}, while optical spectroscopy with an Integral Field Unit is needed to properly constrain the redshifts of the handful of galaxies identified in our optical images to high precision. This will allow us to not only constrain the metal abundances for all intervening systems, but also determine the kinematic properties of the \hi with higher precision (inflow, outflow, rotation), and spatially disentangle all the multiphase components of the gas haloes intersecting \pks \citep[e.g.][]{Peroux2019, Weng2022}. In short, higher spatial and spectral resolution optical data is crucial to further disentangle the light of \pks and the intervening systems, in order to better understand the properties of the gas probed by this not-so-compact radio quasar.

\begin{acknowledgement}
The authors wish to thank Prof. Max Pettini for enlightening discussion and advice on the analysis of DLA absorption systems, Prof. Tom Oosterloo for pointing out data in the the ALMA calibrator catalogue, Dr. Mark Walker for helpful comments on the history surrounding \pksx, and Dr. Kimberly Emig for her helpful comments on a mature version of this manuscript. The authors also wish to thank the anonymous referee for their helpful comments, which improved the overall clarity of this work.

EFK is supported by an Australian Government Research Training Program (RTP) Scholarship.\footnote{doi.org/10.82133/C42F-K220} HY is supported  by the National Research Foundation of Korea (NRF) grant funded by the Korea government (MSIT) (RS-2025-00516062). MG is supported through UK STFC Grant ST/Y001117/1. MG acknowledges support from the Inter-University Institute for Data Intensive Astronomy (IDIA). IDIA is a partnership of the University of Cape Town, the University of Pretoria and the University of the Western Cape. For the purpose of open access, the author has applied a Creative Commons Attribution (CC BY) licence to any Author Accepted Manuscript version arising from this submission

This scientific work uses data obtained from Inyarrimanha Ilgari Bundara, the CSIRO Murchison Radio-astronomy Observatory. We acknowledge the Wajarri Yamaji People as the Traditional Owners and native title holders of the Observatory site. CSIRO’s ASKAP radio telescope is part of the Australia Telescope National Facility (https://ror.org/05qajvd42). Operation of ASKAP is funded by the Australian Government with support from the National Collaborative Research Infrastructure Strategy. ASKAP uses the resources of the Pawsey Supercomputing Research Centre. Establishment of ASKAP, Inyarrimanha Ilgari Bundara, the CSIRO Murchison Radio-astronomy Observatory and the Pawsey Supercomputing Research Centre are initiatives of the Australian Government, with support from the Government of Western Australia and the Science and Industry Endowment Fund.

This paper includes archived data obtained through the CSIRO ASKAP Science Data Archive, CASDA.

Analysis in this paper is based on observations obtained under project GS-2024B-FT-215 (P.I.\ Yoon) at the international Gemini Observatory, a program of NSF NOIRLab, which is managed by the Association of Universities for Research in Astronomy (AURA) under a cooperative agreement with the U.S. National Science Foundation on behalf of the Gemini Observatory partnership: the U.S. National Science Foundation (United States), National Research Council (Canada), Agencia Nacional de Investigaci\'{o}n y Desarrollo (Chile), Ministerio de Ciencia, Tecnolog\'{i}a e Innovaci\'{o}n (Argentina), Minist\'{e}rio da Ci\^{e}ncia, Tecnologia, Inova\c{c}\~{o}es e Comunica\c{c}\~{o}es (Brazil), and Korea Astronomy and Space Science Institute (Republic of Korea).

This research uses services or data provided by the Astro Data Lab, which is part of the Community Science and Data Center (CSDC) Program of NSF NOIRLab. NOIRLab is operated by the Association of Universities for Research in Astronomy (AURA), Inc. under a cooperative agreement with the U.S. National Science Foundation. \citep{Fitzpatrick2014, Nikutta2020, Juneau2021}.

The DESI Legacy Imaging Surveys consist of three individual and complementary projects: the Dark Energy Camera Legacy Survey (DECaLS), the Beijing-Arizona Sky Survey (BASS), and the Mayall z-band Legacy Survey (MzLS). DECaLS, BASS and MzLS together include data obtained, respectively, at the Blanco telescope, Cerro Tololo Inter-American Observatory, NSF’s NOIRLab; the Bok telescope, Steward Observatory, University of Arizona; and the Mayall telescope, Kitt Peak National Observatory, NOIRLab. NOIRLab is operated by the Association of Universities for Research in Astronomy (AURA) under a cooperative agreement with the National Science Foundation. Pipeline processing and analyses of the data were supported by NOIRLab and the Lawrence Berkeley National Laboratory (LBNL). Legacy Surveys also uses data products from the Near-Earth Object Wide-field Infrared Survey Explorer (NEOWISE), a project of the Jet Propulsion Laboratory/California Institute of Technology, funded by the National Aeronautics and Space Administration. Legacy Surveys was supported by: the Director, Office of Science, Office of High Energy Physics of the U.S. Department of Energy; the National Energy Research Scientific Computing Center, a DOE Office of Science User Facility; the U.S. National Science Foundation, Division of Astronomical Sciences; the National Astronomical Observatories of China, the Chinese Academy of Sciences and the Chinese National Natural Science Foundation. LBNL is managed by the Regents of the University of California under contract to the U.S. Department of Energy. The complete acknowledgments can be found at https://www.legacysurvey.org/acknowledgment/.

The Photometric Redshifts for the Legacy Surveys (PRLS) catalogue used in this paper was produced thanks to funding from the U.S. Department of Energy Office of Science, Office of High Energy Physics via grant DE-SC0007914.

\end{acknowledgement}


\bibliography{pks0405-bib}

@article{Kanekar2001,
    title = {{Variable 21-cm absorption at z=0.3127}},
    year = {2001},
    journal = {Monthly Notices of the Royal Astronomical Society},
    author = {Kanekar, N. and Chengalur, J. N.},
    number = {2},
    month = {8},
    pages = {631--635},
    volume = {325},
    doi = {10.1046/j.1365-8711.2001.04424.x},
    issn = {0035-8711}
}

@article{Kedziora-Chudczer1997,
    title = {{PKS 0405−385: The Smallest Radio Quasar?}},
    year = {1997},
    journal = {The Astrophysical Journal},
    author = {Kedziora-Chudczer, L. and Jauncey, D. L. and Wieringa, M. H. and Walker, M. A. and Nicolson, G. D. and Reynolds, J. E. and Tzioumis, A. K.},
    number = {1},
    month = {11},
    pages = {L9-L12},
    volume = {490},
    doi = {10.1086/311001},
    issn = {0004637X}
}

@article{Gong2022,
    title = {{Quasiperiodic Behavior in the {$\gamma$}-Ray Light Curve of the Blazar PKS 0405-385}},
    year = {2022},
    journal = {The Astrophysical Journal},
    author = {Gong, Yunlu and Zhou, Liancheng and Yuan, Min and Zhang, Haiyun and Yi, Tingfeng and Fang, Jun},
    number = {2},
    month = {6},
    pages = {168},
    volume = {931},
    doi = {10.3847/1538-4357/ac6c8c},
    issn = {0004-637X}
}

@article{Charlot2020,
    title = {{The third realization of the International Celestial Reference Frame by very long baseline interferometry}},
    year = {2020},
    journal = {Astronomy {\&} Astrophysics},
    author = {Charlot, P. and Jacobs, C. S. and Gordon, D. and Lambert, S. and de Witt, A. and B{\"{o}}hm, J. and Fey, A. L. and Heinkelmann, R. and Skurikhina, E. and Titov, O. and Arias, E. F. and Bolotin, S. and Bourda, G. and Ma, C. and Malkin, Z. and Nothnagel, A. and Mayer, D. and MacMillan, D. S. and Nilsson, T. and Gaume, R.},
    month = {12},
    pages = {A159},
    volume = {644},
    doi = {10.1051/0004-6361/202038368},
    issn = {0004-6361}
}

@article{Veron1990,
    title = {{Miscellaneous spectroscopic observations of quasars and quasar candidates.}},
    year = {1990},
    author = {V{\'e}ron, P and V{\'e}ron-Cetty, M -P. and Djorgovski, S and Magain, P and Meylan, G and Surdej, J},
    journal = {Astronomy and Astrophysics Supplement},
    month = {12},
    pages = {543},
    volume = {86}
}

@article{Savage1981,
    title = {{Identification of southern radio sources - IV}},
    year = {1981},
    journal = {Monthly Notices of the Royal Astronomical Society},
    author = {Savage, A. and Wright, A. E.},
    number = {4},
    month = {10},
    pages = {927--932},
    volume = {196},
    doi = {10.1093/mnras/196.4.927},
    issn = {0035-8711}
}

@article{Yoon2025,
    title = {{The first large absorption survey in H <scp>i</scp> (FLASH): II. Pilot survey data release and first results}},
    year = {2025},
    journal = {Publications of the Astronomical Society of Australia},
    author = {Yoon, Hyein and Sadler, Elaine M. and Mahony, Elizabeth K. and Aditya, J.N.H.S. and Allison, James R. and Glowacki, Marcin and Kerrison, Emily F. and Moss, Vanessa A. and Su, Renzhi and Weng, Simon and Whiting, Matthew and Wong, O. Ivy and Callingham, Joseph R. and Curran, Stephen J. and Darling, Jeremy and Edge, Alastair C. and Ellison, Sara L. and Emig, Kimberly L. and Garratt-Smithson, Lilian and German, Gordon and Grasha, Kathryn and Koribalski, Bärbel S. and Morganti, Raffaella and Oosterloo, Tom and P{\'{e}}roux, Céline and Pettini, Max and Pimbblet, Kevin A. and Zheng, Zheng and Zwaan, Martin and Ball, Lewis and Bock, Douglas C.-J. and Brodrick, David and Bunton, John D. and Cooray, F.R. and Edwards, Philip G. and Hayman, Douglas B. and Hotan, Aidan W. and Lee-Waddell, K. and McClure-Griffiths, N.M. and Ng, A. and Phillips, Chris J. and Raja, Wasim and Voronkov, Maxim A. and Westmeier, Tobias},
    month = {6},
    pages = {e088},
    volume = {42},
    doi = {10.1017/pasa.2025.10046},
    issn = {1323-3580}
}

@article{Allison2022,
    title = {{The First Large Absorption Survey in HI (FLASH): I. Science Goals and Survey Design}},
    year = {2022},
    journal = {PASA},
    author = {Allison, J. R. and Sadler, E. M. and Amaral, A. D. and An, T. and Curran, S. J. and Darling, J. and Edge, A. C. and Ellison, S. L. and Emig, K. L. and Gaensler, B. M. and Garratt-Smithson, L. and Glowacki, M. and Grasha, K. and Koribalski, B. S. and Lagos, C. del P. and Lah, P. and Mahony, E. K. and Mao, S. A. and Morganti, R. and Moss, V. A. and Pettini, M. and Pimbblet, K. A. and Power, C. and Salas, P. and Staveley-Smith, L. and Whiting, M. T. and Wong, O. I. and Yoon, H. and Zheng, Z. and Zwaan, M. A.},
    number = {},
    month = {10},
    pages = {010--1},
    volume = {39},
    url = {http://arxiv.org/abs/2110.00469},
    arxivId = {2110.00469}
}

@article{Hotan2021,
    title = {{Australian square kilometre array pathfinder: I. system description}},
    year = {2021},
    journal = {PASA},
    author = {Hotan, A. W. and Bunton, J. D. and Chippendale, A. P. and Whiting, M. and Tuthill, J. and Moss, V. A. and McConnell, D. and Amy, S. W. and Huynh, M. T. and Allison, J. R. and Anderson, C. S. and Bannister, K. W. and Bastholm, E. and Beresford, R. and Bock, D. C. -J. and Bolton, R. and Chapman, J. M. and Chow, K. and Collier, J. D. and Cooray, F. R. and Cornwell, T. J. and Diamond, P. J. and Edwards, P. G. and Feain, I. J. and Franzen, T. M. O. and George, D. and Gupta, N. and Hampson, G. A. and Harvey-Smith, L. and Hayman, D. B. and Heywood, I. and Jacka, C. and Jackson, C. A. and Jackson, S. and Jeganathan, K. and Johnston, S. and Kesteven, M. and Kleiner, D. and Koribalski, B. S. and Lee-Waddell, K. and Lenc, E. and Lensson, E. S. and Mackay, S. and Mahony, E. K. and McClure-Griffiths, N. M. and McConigley, R. and Mirtschin, P. and Ng, A. K. and Norris, R. P. and Pearce, S. E. and Phillips, C. and Pilawa, M. A. and Raja, W. and Reynolds, J. E. and Roberts, P. and Roxby, D. N. and Sadler, E. M. and Shields, M. and Schinckel, A. E. T. and Serra, P. and Shaw, R. D. and Sweetnam, T. and Troup, E. R. and Tzioumis, A. and Voronkov, M. A. and Westmeier, T.},
    month = {3},
    pages = {e009},
    volume = {38},
    publisher = {Cambridge University Press (CUP)},
    url = {https://ui.adsabs.harvard.edu/abs/2021PASA...38....9H/abstract},
    doi = {10.1017/PASA.2021.1},
    issn = {1323-3580}
}

@article{Allison2012,
    title = {{Application of a Bayesian Method to Absorption Spectral-Line Finding in Simulated ASKAP Data}},
    year = {2012},
    journal = {PASA},
    author = {Allison, J. R. and Sadler, E. M. and Whiting, M. T.},
    number = {3},
    pages = {221--228},
    volume = {29},
    url = {https://ui.adsabs.harvard.edu/abs/2012PASA...29..221A/abstract},
    doi = {10.1071/AS11040},
    issn = {1323-3580},
    arxivId = {arXiv:1109.3539}
}

@article{Kanekar2003,
    title = {{HI absorption in a gravitational lens at <i>z</i> {\~{}} 0.7645}},
    year = {2003},
    journal = {Astronomy {\&} Astrophysics},
    author = {Kanekar, N. and Briggs, F. H.},
    number = {2},
    month = {12},
    pages = {L29-L32},
    volume = {412},
    doi = {10.1051/0004-6361:20031676},
    issn = {0004-6361}
}

@article{Labrie2023,
    title = {{DRAGONS—A Quick Overview}},
    year = {2023},
    journal = {Research Notes of the AAS},
    author = {Labrie, K. and Simpson, C. and Cardenes, R. and Turner, J. and Soraisam, M. and Quint, B. and Oberdorf, O. and Placco, V. M. and Berke, D. and Smirnova, O. and Conseil, S. and Vacca, W. D. and Thomas-Osip, J.},
    number = {10},
    month = {10},
    pages = {214},
    volume = {7},
    doi = {10.3847/2515-5172/ad0044},
    issn = {2515-5172}
}

@article{Hinton2016,
    title = {{Marz: Manual and automatic redshifting software}},
    year = {2016},
    journal = {Astronomy and Computing},
    author = {Hinton, S.R. and Davis, Tamara M. and Lidman, C. and Glazebrook, K. and Lewis, G.F.},
    month = {4},
    pages = {61--71},
    volume = {15},
    doi = {10.1016/j.ascom.2016.03.001},
    issn = {22131337}
}

@ARTICLE{wolfe05,
   author = {{Wolfe}, A.~M. and {Gawiser}, E. and {Prochaska}, J.~X.},
    title = "{Damped Ly {$\alpha$} Systems}",
  journal = {\araa},
   eprint = {astro-ph/0509481},
     year = 2005,
    month = sep,
   volume = 43,
    pages = {861-918},
      doi = {10.1146/annurev.astro.42.053102.133950},
   adsurl = {http://cdsads.u-strasbg.fr/abs/2005ARA\%26A..43..861W}
}

@ARTICLE{kanekar04,
   author = {{Kanekar}, N. and {Briggs}, F.~H.},
    title = "{21-cm absorption studies with the Square Kilometer Array}",
  journal = {\nar},
   eprint = {astro-ph/0409169},
     year = 2004,
    month = dec,
   volume = 48,
    pages = {1259-1270},
      doi = {10.1016/j.newar.2004.09.030},
   adsurl = {http://adsabs.harvard.edu/abs/2004NewAR..48.1259K}
}

@ARTICLE{morganti15,
   author = {{Morganti}, R. and {Sadler}, E.~M. and {Curran}, S.},
    title = "{Cool Outflows and HI absorbers with SKA}",
  journal = {Advancing Astrophysics with the Square Kilometre Array (AASKA14)},
archivePrefix = "arXiv",
   eprint = {1501.01091},
     year = 2015,
    month = apr,
      eid = {134},
    pages = {134},
   adsurl = {http://adsabs.harvard.edu/abs/2015aska.confE.134M}
}

@article{Curran2021,
    title = {{Intervening or associated? Machine learning classification of redshifted H I 21-cm absorption}},
    year = {2021},
    journal = {MNRAS},
    author = {Curran, S J.},
    number = {1},
    month = {7},
    pages = {1548--1556},
    volume = {506},
    url = {https://ui.adsabs.harvard.edu/abs/2021MNRAS.506.1548C/abstract},
    doi = {10.1093/MNRAS/STAB1865},
    issn = {0035-8711},
    arxivId = {arXiv:2106.14551}
}

@article{Juneau2021,
    title = {{Jupyter-Enabled Astrophysical Analysis Using Data-Proximate Computing Platforms}},
    year = {2021},
    journal = {Computing in Science {\&} Engineering},
    author = {Juneau, Stéphanie and Olsen, Knut and Nikutta, Robert and Jacques, Alice and Bailey, Stephen},
    number = {2},
    pages = {15--25},
    volume = {23},
    doi = {10.1109/MCSE.2021.3057097}
}

@article{Nikutta2020,
    title = {{Data Lab—A community science platform}},
    year = {2020},
    journal = {Astronomy and Computing},
    author = {Nikutta, R and Fitzpatrick, M and Scott, A and Weaver, B A},
    pages = {100411},
    volume = {33},
    url = {https://www.sciencedirect.com/science/article/pii/S2213133720300652},
    doi = {https://doi.org/10.1016/j.ascom.2020.100411},
    issn = {2213-1337}
}

@inproceedings{Fitzpatrick2014,
    title = {{The NOAO Data Laboratory: a conceptual overview}},
    year = {2014},
    booktitle = {Observatory Operations: Strategies, Processes, and Systems V},
    author = {Fitzpatrick, Michael J and Olsen, Knut and Economou, Frossie and Stobie, Elizabeth B and Beers, T.~C. and Dickinson, Mark and Norris, Patrick and Saha, Abi and Seaman, Robert and Silva, David R and Swaters, Robert A and Thomas, Brian and Valdes, Francisco},
    editor = {Peck, Alison B and Benn, Chris R and Seaman, Robert L},
    month = {8},
    pages = {91491T},
    series = {Society of Photo-Optical Instrumentation Engineers (SPIE) Conference Series},
    volume = {9149},
    doi = {10.1117/12.2057445}
}

@article{Dutta2017,
    title = {{Incidence of Hi 21-cm absorption in strong Feii systems at 0.5 {\&}lt; <i>z</i> {\&}lt; 1.5}},
    year = {2017},
    journal = {Monthly Notices of the Royal Astronomical Society},
    author = {Dutta, R. and Srianand, R. and Gupta, N. and Joshi, R. and Petitjean, P. and Noterdaeme, P. and Ge, J. and Krogager, J.-K.},
    number = {4},
    month = {3},
    pages = {4249--4264},
    volume = {465},
    doi = {10.1093/mnras/stw3040},
    issn = {0035-8711}
}

@misc{Specutils,
    title = {{Specutils: Spectroscopic analysis and reduction}},
    year = {2019},
    author = {{Astropy-Specutils Development Team}},
    month = {2},
    howpublished = {Astrophysics Source Code Library, record ascl:1902.012}
}

@article{Planck2018,
    title = {{<i>Planck</i> 2018 results}},
    year = {2020},
    journal = {Astronomy {\&} Astrophysics},
    author = {{Planck Collaboration}},
    month = {9},
    pages = {A6},
    volume = {641},
    doi = {10.1051/0004-6361/201833910},
    issn = {0004-6361}
}

@article{Rickett2002,
    title = {{The Role of Interstellar Scintillation in Intraday Variations at Centimetre Wavelengths}},
    year = {2002},
    journal = {Publications of the Astronomical Society of Australia},
    author = {Rickett, Barney},
    number = {1},
    month = {3},
    pages = {100--105},
    volume = {19},
    doi = {10.1071/AS01120},
    issn = {1323-3580}
}

@article{Macquart2005,
    title = {{Scintillation-induced variability in radio absorption spectra against extragalactic sources}},
    year = {2005},
    journal = {Astronomy {\&} Astrophysics},
    author = {Macquart, J.-P.},
    number = {3},
    month = {4},
    pages = {827--840},
    volume = {433},
    doi = {10.1051/0004-6361:20034425},
    issn = {0004-6361}
}

@INPROCEEDINGS{Lucyna2001,
       author = {{Kedziora-Chudczer}, L. and {Jauncey}, D.~L. and {Lovell}, J.~E.~J. and {Walker}, M.~A. and {Macquart}, J. -P. and {Wieringa}, M.~H. and {Tzioumis}, A.~K. and {Perley}, R.~A. and {Reynolds}, J.~E.},
        title = "{Long term monitoring of the extreme intraday variable quasar PKS 0405-385}",
    booktitle = {Particles and Fields in Radio Galaxies Conference},
         year = 2001,
       editor = {{Laing}, Robert A. and {Blundell}, Katherine M.},
       series = {Astronomical Society of the Pacific Conference Series},
       volume = {250},
        month = jan,
        pages = {128},
       adsurl = {https://ui.adsabs.harvard.edu/abs/2001ASPC..250..128K}
}

@ARTICLE{Stevens2012,
       author = {{Stevens}, J. and {Edwards}, P.~G. and {Ojha}, R. and {Kadler}, M. and {Hungwe}, F. and {Dutka}, M. and {Tingay}, S. and {Macquart}, J.~P. and {Moin}, A. and {Lovell}, J. and {Blanchard}, J.},
        title = "{ATCA monitoring of gamma-ray loud AGN}",
      journal = {arXiv e-prints},
         year = 2012,
        month = may,
          eid = {arXiv:1205.2403},
        pages = {arXiv:1205.2403},
          doi = {10.48550/arXiv.1205.2403},
archivePrefix = {arXiv},
       eprint = {1205.2403},
 primaryClass = {astro-ph.HE},
       adsurl = {https://ui.adsabs.harvard.edu/abs/2012arXiv1205.2403S}
}

@ARTICLE{2005AJ....129.1993M,
       author = {{Manchester}, R.~N. and {Hobbs}, G.~B. and {Teoh}, A. and {Hobbs}, M.},
        title = "{The Australia Telescope National Facility Pulsar Catalogue}",
      journal = {\aj},
         year = 2005,
        month = apr,
       volume = {129},
       number = {4},
        pages = {1993-2006},
          doi = {10.1086/428488},
archivePrefix = {arXiv},
       eprint = {astro-ph/0412641},
 primaryClass = {astro-ph},
       adsurl = {https://ui.adsabs.harvard.edu/abs/2005AJ....129.1993M}
}

@ARTICLE{Aditya2025,
       author = {{Aditya}, J.~N.~H.~S. and {Sadler}, Elaine M. and {Morganti}, Raffaella and {Petrov}, L.~Y. and {Tao}, An and {Kerrison}, Emily F. and {Mahony}, Elizabeth K. and {Yoon}, Hyein and {Su}, Renzhi and {Whiting}, Matthew and {Moss}, Vanessa A. and {Weng}, Simon},
        title = "{VLBI studies of FLASH H I 21-cm absorbers -- I}",
      journal = {arXiv e-prints},
         year = 2025,
        month = jul,
          eid = {arXiv:2507.19957},
        pages = {arXiv:2507.19957},
          doi = {10.48550/arXiv.2507.19957},
archivePrefix = {arXiv},
       eprint = {2507.19957},
 primaryClass = {astro-ph.GA},
       adsurl = {https://ui.adsabs.harvard.edu/abs/2025arXiv250719957A}
}

@ARTICLE{Dey2019,
       author = {{Dey}, Arjun and {Schlegel}, David J. and {Lang}, Dustin and {Blum}, Robert and {Burleigh}, Kaylan and {Fan}, Xiaohui and {Findlay}, Joseph R. and {Finkbeiner}, Doug and {Herrera}, David and {Juneau}, St{\'e}phanie and {Landriau}, Martin and {Levi}, Michael and {McGreer}, Ian and {Meisner}, Aaron and {Myers}, Adam D. and {Moustakas}, John and {Nugent}, Peter and {Patej}, Anna and {Schlafly}, Edward F. and {Walker}, Alistair R. and {Valdes}, Francisco and {Weaver}, Benjamin A. and {Y{\`e}che}, Christophe and {Zou}, Hu and {Zhou}, Xu and {Abareshi}, Behzad and {Abbott}, T.~M.~C. and {Abolfathi}, Bela and {Aguilera}, C. and {Alam}, Shadab and {Allen}, Lori and {Alvarez}, A. and {Annis}, James and {Ansarinejad}, Behzad and {Aubert}, Marie and {Beechert}, Jacqueline and {Bell}, Eric F. and {BenZvi}, Segev Y. and {Beutler}, Florian and {Bielby}, Richard M. and {Bolton}, Adam S. and {Brice{\~n}o}, C{\'e}sar and {Buckley-Geer}, Elizabeth J. and {Butler}, Karen and {Calamida}, Annalisa and {Carlberg}, Raymond G. and {Carter}, Paul and {Casas}, Ricard and {Castander}, Francisco J. and {Choi}, Yumi and {Comparat}, Johan and {Cukanovaite}, Elena and {Delubac}, Timoth{\'e}e and {DeVries}, Kaitlin and {Dey}, Sharmila and {Dhungana}, Govinda and {Dickinson}, Mark and {Ding}, Zhejie and {Donaldson}, John B. and {Duan}, Yutong and {Duckworth}, Christopher J. and {Eftekharzadeh}, Sarah and {Eisenstein}, Daniel J. and {Etourneau}, Thomas and {Fagrelius}, Parker A. and {Farihi}, Jay and {Fitzpatrick}, Mike and {Font-Ribera}, Andreu and {Fulmer}, Leah and {G{\"a}nsicke}, Boris T. and {Gaztanaga}, Enrique and {George}, Koshy and {Gerdes}, David W. and {Gontcho}, Satya Gontcho A. and {Gorgoni}, Claudio and {Green}, Gregory and {Guy}, Julien and {Harmer}, Diane and {Hernandez}, M. and {Honscheid}, Klaus and {Huang}, Lijuan Wendy and {James}, David J. and {Jannuzi}, Buell T. and {Jiang}, Linhua and {Joyce}, Richard and {Karcher}, Armin and {Karkar}, Sonia and {Kehoe}, Robert and {Kneib}, Jean-Paul and {Kueter-Young}, Andrea and {Lan}, Ting-Wen and {Lauer}, Tod R. and {Le Guillou}, Laurent and {Le Van Suu}, Auguste and {Lee}, Jae Hyeon and {Lesser}, Michael and {Perreault Levasseur}, Laurence and {Li}, Ting S. and {Mann}, Justin L. and {Marshall}, Robert and {Mart{\'\i}nez-V{\'a}zquez}, C.~E. and {Martini}, Paul and {du Mas des Bourboux}, H{\'e}lion and {McManus}, Sean and {Meier}, Tobias Gabriel and {M{\'e}nard}, Brice and {Metcalfe}, Nigel and {Mu{\~n}oz-Guti{\'e}rrez}, Andrea and {Najita}, Joan and {Napier}, Kevin and {Narayan}, Gautham and {Newman}, Jeffrey A. and {Nie}, Jundan and {Nord}, Brian and {Norman}, Dara J. and {Olsen}, Knut A.~G. and {Paat}, Anthony and {Palanque-Delabrouille}, Nathalie and {Peng}, Xiyan and {Poppett}, Claire L. and {Poremba}, Megan R. and {Prakash}, Abhishek and {Rabinowitz}, David and {Raichoor}, Anand and {Rezaie}, Mehdi and {Robertson}, A.~N. and {Roe}, Natalie A. and {Ross}, Ashley J. and {Ross}, Nicholas P. and {Rudnick}, Gregory and {Safonova}, Sasha and {Saha}, Abhijit and {S{\'a}nchez}, F. Javier and {Savary}, Elodie and {Schweiker}, Heidi and {Scott}, Adam and {Seo}, Hee-Jong and {Shan}, Huanyuan and {Silva}, David R. and {Slepian}, Zachary and {Soto}, Christian and {Sprayberry}, David and {Staten}, Ryan and {Stillman}, Coley M. and {Stupak}, Robert J. and {Summers}, David L. and {Sien Tie}, Suk and {Tirado}, H. and {Vargas-Maga{\~n}a}, Mariana and {Vivas}, A. Katherina and {Wechsler}, Risa H. and {Williams}, Doug and {Yang}, Jinyi and {Yang}, Qian and {Yapici}, Tolga and {Zaritsky}, Dennis and {Zenteno}, A. and {Zhang}, Kai and {Zhang}, Tianmeng and {Zhou}, Rongpu and {Zhou}, Zhimin},
        title = "{Overview of the DESI Legacy Imaging Surveys}",
      journal = {\aj},
         year = 2019,
        month = may,
       volume = {157},
       number = {5},
          eid = {168},
        pages = {168},
          doi = {10.3847/1538-3881/ab089d},
archivePrefix = {arXiv},
       eprint = {1804.08657},
 primaryClass = {astro-ph.IM},
       adsurl = {https://ui.adsabs.harvard.edu/abs/2019AJ....157..168D}
}

@ARTICLE{Adelberger2005,
       author = {{Adelberger}, Kurt L. and {Shapley}, Alice E. and {Steidel}, Charles C. and {Pettini}, Max and {Erb}, Dawn K. and {Reddy}, Naveen A.},
        title = "{The Connection between Galaxies and Intergalactic Absorption Lines at Redshift 2<\raisebox{-0.5ex}\textasciitildez<\raisebox{-0.5ex}\textasciitilde3}",
      journal = {\apj},
         year = 2005,
        month = aug,
       volume = {629},
       number = {2},
        pages = {636-653},
          doi = {10.1086/431753},
archivePrefix = {arXiv},
       eprint = {astro-ph/0505122},
 primaryClass = {astro-ph},
       adsurl = {https://ui.adsabs.harvard.edu/abs/2005ApJ...629..636A}
}

@ARTICLE{Weng2022,
       author = {{Weng}, Simon and {Sadler}, Elaine M. and {Foster}, Caroline and {P{\'e}roux}, C{\'e}line and {Mahony}, Elizabeth K. and {Allison}, James R. and {Moss}, Vanessa A. and {Su}, Renzhi and {Whiting}, Matthew T. and {Yoon}, Hyein},
        title = "{Observations of cold extragalactic gas clouds at z = 0.45 towards PKS 1610-771}",
      journal = {\mnras},
         year = 2022,
        month = may,
       volume = {512},
       number = {3},
        pages = {3638-3650},
          doi = {10.1093/mnras/stac747},
archivePrefix = {arXiv},
       eprint = {2109.10875},
 primaryClass = {astro-ph.GA},
       adsurl = {https://ui.adsabs.harvard.edu/abs/2022MNRAS.512.3638W}
}

@ARTICLE{Madau2014,
       author = {{Madau}, Piero and {Dickinson}, Mark},
        title = "{Cosmic Star-Formation History}",
      journal = {\araa},
         year = 2014,
        month = aug,
       volume = {52},
        pages = {415-486},
          doi = {10.1146/annurev-astro-081811-125615},
archivePrefix = {arXiv},
       eprint = {1403.0007},
 primaryClass = {astro-ph.CO},
       adsurl = {https://ui.adsabs.harvard.edu/abs/2014ARA&A..52..415M}
}

@ARTICLE{Norris2011,
       author = {{Norris}, Ray P. and {Hopkins}, A.~M. and {Afonso}, J. and {Brown}, S. and {Condon}, J.~J. and {Dunne}, L. and {Feain}, I. and {Hollow}, R. and {Jarvis}, M. and {Johnston-Hollitt}, M. and {Lenc}, E. and {Middelberg}, E. and {Padovani}, P. and {Prandoni}, I. and {Rudnick}, L. and {Seymour}, N. and {Umana}, G. and {Andernach}, H. and {Alexander}, D.~M. and {Appleton}, P.~N. and {Bacon}, D. and {Banfield}, J. and {Becker}, W. and {Brown}, M.~J.~I. and {Ciliegi}, P. and {Jackson}, C. and {Eales}, S. and {Edge}, A.~C. and {Gaensler}, B.~M. and {Giovannini}, G. and {Hales}, C.~A. and {Hancock}, P. and {Huynh}, M.~T. and {Ibar}, E. and {Ivison}, R.~J. and {Kennicutt}, R. and {Kimball}, Amy E. and {Koekemoer}, A.~M. and {Koribalski}, B.~S. and {L{\'o}pez-S{\'a}nchez}, {\'A}. R. and {Mao}, M.~Y. and {Murphy}, T. and {Messias}, H. and {Pimbblet}, K.~A. and {Raccanelli}, A. and {Randall}, K.~E. and {Reiprich}, T.~H. and {Roseboom}, I.~G. and {R{\"o}ttgering}, H. and {Saikia}, D.~J. and {Sharp}, R.~G. and {Slee}, O.~B. and {Smail}, Ian and {Thompson}, M.~A. and {Urquhart}, J.~S. and {Wall}, J.~V. and {Zhao}, G.-B.},
        title = "{EMU: Evolutionary Map of the Universe}",
      journal = {\pasa},
         year = 2011,
        month = aug,
       volume = {28},
       number = {3},
        pages = {215-248},
          doi = {10.1071/AS11021},
archivePrefix = {arXiv},
       eprint = {1106.3219},
 primaryClass = {astro-ph.CO},
       adsurl = {https://ui.adsabs.harvard.edu/abs/2011PASA...28..215N}
}

@article{Rickett2002b, 
    title={Interstellar Scintillation Explains the Intraday Variations in the Linear Polarisation of Quasar PKS 0405-385 at cm-wavelengths}, 
    volume={19}, DOI={10.1071/AS01119}, 
    number={1}, 
    journal={Publications of the Astronomical Society of Australia}, 
    author={Rickett, Barney and Kedziora-Chudczer, Lucyna and Jauncey, David L.}, 
    year={2002}, 
    pages={106–110}}

@ARTICLE{Kalberla2009,
       author = {{Kalberla}, Peter M.~W. and {Kerp}, J{\"u}rgen},
        title = "{The Hi Distribution of the Milky Way}",
      journal = {\araa},
         year = 2009,
        month = sep,
       volume = {47},
       number = {1},
        pages = {27-61},
          doi = {10.1146/annurev-astro-082708-101823},
       adsurl = {https://ui.adsabs.harvard.edu/abs/2009ARA&A..47...27K}
}

@ARTICLE{Abdollahi2023,
       author = {{Abdollahi}, S. and {Ajello}, M. and {Baldini}, L. and {Ballet}, J. and {Bastieri}, D. and {Becerra Gonzalez}, J. and {Bellazzini}, R. and {Berretta}, A. and {Bissaldi}, E. and {Bonino}, R. and {Brill}, A. and {Bruel}, P. and {Burns}, E. and {Buson}, S. and {Cameron}, R.~A. and {Caputo}, R. and {Caraveo}, P.~A. and {Cibrario}, N. and {Ciprini}, S. and {Cristarella Orestano}, P. and {Crnogorcevic}, M. and {Cutini}, S. and {D'Ammando}, F. and {De Gaetano}, S. and {Digel}, S.~W. and {Di Lalla}, N. and {Di Venere}, L. and {Dom{\'\i}nguez}, A. and {Ramazani}, V. Fallah and {Fegan}, S.~J. and {Ferrara}, E.~C. and {Fiori}, A. and {Fleischhack}, H. and {Franckowiak}, A. and {Fukazawa}, Y. and {Fusco}, P. and {Gammaldi}, V. and {Gargano}, F. and {Garrappa}, S. and {Gasbarra}, C. and {Gasparrini}, D. and {Giglietto}, N. and {Giordano}, F. and {Giroletti}, M. and {Green}, D. and {Grenier}, I.~A. and {Guiriec}, S. and {Gustafsson}, M. and {Hays}, E. and {Horan}, D. and {Hou}, X. and {J{\'o}hannesson}, G. and {Kerr}, M. and {Kocevski}, D. and {Kuss}, M. and {Latronico}, L. and {Li}, J. and {Liodakis}, I. and {Longo}, F. and {Loparco}, F. and {Lorusso}, L. and {Lott}, B. and {Lovellette}, M.~N. and {Lubrano}, P. and {Maldera}, S. and {Manfreda}, A. and {Mart{\'\i}-Devesa}, G. and {Mazziotta}, M.~N. and {Mereu}, I. and {Meyer}, M. and {Michelson}, P.~F. and {Mizuno}, T. and {Monzani}, M.~E. and {Morselli}, A. and {Moskalenko}, I.~V. and {Negro}, M. and {Omodei}, N. and {Orlando}, E. and {Ormes}, J.~F. and {Paneque}, D. and {Panzarini}, G. and {Perkins}, J.~S. and {Persic}, M. and {Pesce-Rollins}, M. and {Pillera}, R. and {Porter}, T.~A. and {Principe}, G. and {Racusin}, J.~L. and {Rain{\`o}}, S. and {Rando}, R. and {Rani}, B. and {Razzano}, M. and {Razzaque}, S. and {Reimer}, A. and {Reimer}, O. and {S{\'a}nchez-Conde}, M. and {Parkinson}, P.~M. Saz and {Scargle}, Jeff and {Scotton}, L. and {Serini}, D. and {Sgr{\`o}}, C. and {Siskind}, E.~J. and {Spandre}, G. and {Spinelli}, P. and {Suson}, D.~J. and {Tajima}, H. and {Thompson}, D.~J. and {Torres}, D.~F. and {Valverde}, J. and {Venters}, T. and {Wadiasingh}, Z. and {Wagner}, S. and {Wood}, K.},
        title = "{The Fermi-LAT Lightcurve Repository}",
      journal = {\apjs},
         year = 2023,
        month = apr,
       volume = {265},
       number = {2},
          eid = {31},
        pages = {31},
          doi = {10.3847/1538-4365/acbb6a},
archivePrefix = {arXiv},
       eprint = {2301.01607},
 primaryClass = {astro-ph.HE},
       adsurl = {https://ui.adsabs.harvard.edu/abs/2023ApJS..265...31A}
}

@ARTICLE{McKee2015,
       author = {{McKee}, Christopher F. and {Parravano}, Antonio and {Hollenbach}, David J.},
        title = "{Stars, Gas, and Dark Matter in the Solar Neighborhood}",
      journal = {\apj},
         year = 2015,
        month = nov,
       volume = {814},
       number = {1},
          eid = {13},
        pages = {13},
          doi = {10.1088/0004-637X/814/1/13},
archivePrefix = {arXiv},
       eprint = {1509.05334},
 primaryClass = {astro-ph.GA},
       adsurl = {https://ui.adsabs.harvard.edu/abs/2015ApJ...814...13M}
}

@ARTICLE{2025NatAs...9.1053R,
       author = {{Reardon}, Daniel J. and {Main}, Robert and {Ocker}, Stella Koch and {Shannon}, Ryan M. and {Bailes}, Matthew and {Camilo}, Fernando and {Geyer}, Marisa and {Jameson}, Andrew and {Kramer}, Michael and {Parthasarathy}, Aditya and {Spiewak}, Ren{\'e}e and {van Straten}, Willem and {Venkatraman Krishnan}, Vivek},
        title = "{Bow shock and Local Bubble plasma unveiled by the scintillating millisecond pulsar J0437{\ensuremath{-}}4715}",
      journal = {Nature Astronomy},
         year = 2025,
        month = jul,
       volume = {9},
        pages = {1053-1063},
          doi = {10.1038/s41550-025-02534-6},
archivePrefix = {arXiv},
       eprint = {2410.21390},
 primaryClass = {astro-ph.HE},
       adsurl = {https://ui.adsabs.harvard.edu/abs/2025NatAs...9.1053R}
}

@ARTICLE{2021MNRAS.502.3294W,
       author = {{Wang}, Yuanming and {Tuntsov}, Artem and {Murphy}, Tara and {Lenc}, Emil and {Walker}, Mark and {Bannister}, Keith and {Kaplan}, David L. and {Mahony}, Elizabeth K.},
        title = "{ASKAP observations of multiple rapid scintillators reveal a degrees-long plasma filament}",
      journal = {\mnras},
         year = 2021,
        month = apr,
       volume = {502},
       number = {3},
        pages = {3294-3311},
          doi = {10.1093/mnras/stab139},
archivePrefix = {arXiv},
       eprint = {2101.06048},
 primaryClass = {astro-ph.GA},
       adsurl = {https://ui.adsabs.harvard.edu/abs/2021MNRAS.502.3294W}
}

@software{dstools,
       author = {{Pritchard}, Joshua},
        title = "{askap-vast/dstools: v2.0.0}",
         year = 2025,
        month = apr,
          eid = {10.5281/zenodo.15232974},
          doi = {10.5281/zenodo.15232974},
      version = {v2.0.0},
    publisher = {Zenodo},
       adsurl = {https://ui.adsabs.harvard.edu/abs/2025zndo..15232974P}
}

@ARTICLE{Lucyna2006,
       author = {{Kedziora-Chudczer}, L.},
        title = "{Long-term monitoring of the intra-day variable quasar PKS 0405-385}",
      journal = {\mnras},
         year = 2006,
        month = jun,
       volume = {369},
       number = {1},
        pages = {449-464},
          doi = {10.1111/j.1365-2966.2006.10321.x},
       adsurl = {https://ui.adsabs.harvard.edu/abs/2006MNRAS.369..449K}
}

@ARTICLE{Martinez2018,
       author = {{Mart{\'\i}nez-Aldama}, M.~L. and {del Olmo}, A. and {Marziani}, P. and {Sulentic}, J.~W. and {Negrete}, C.~A. and {Dultzin}, D. and {D'Onofrio}, M. and {Perea}, J.},
        title = "{Extreme quasars at high redshift}",
      journal = {\aap},
         year = 2018,
        month = nov,
       volume = {618},
          eid = {A179},
        pages = {A179},
          doi = {10.1051/0004-6361/201833541},
archivePrefix = {arXiv},
       eprint = {1807.11006},
 primaryClass = {astro-ph.GA},
       adsurl = {https://ui.adsabs.harvard.edu/abs/2018A&A...618A.179M}
}

@article{Marziani2014,
    author = {Marziani, P. and Sulentic, J. W.},
    title = {Highly accreting quasars: sample definition and possible cosmological implications},
    journal = {Monthly Notices of the Royal Astronomical Society},
    volume = {442},
    number = {2},
    pages = {1211-1229},
    year = {2014},
    month = {06},
    issn = {0035-8711},
    doi = {10.1093/mnras/stu951},
    url = {https://doi.org/10.1093/mnras/stu951},
    eprint = {https://academic.oup.com/mnras/article-pdf/442/2/1211/5819740/stu951.pdf},
}

@ARTICLE{Morganti2018,
       author = {{Morganti}, Raffaella and {Oosterloo}, Tom},
        title = "{The interstellar and circumnuclear medium of active nuclei traced by H i 21 cm absorption}",
      journal = {\aapr},
         year = 2018,
        month = jul,
       volume = {26},
       number = {1},
          eid = {4},
        pages = {4},
          doi = {10.1007/s00159-018-0109-x},
archivePrefix = {arXiv},
       eprint = {1807.01475},
 primaryClass = {astro-ph.GA},
       adsurl = {https://ui.adsabs.harvard.edu/abs/2018A&ARv..26....4M}
}

@ARTICLE{allison2017,
       author = {{Allison}, J.~R. and {Moss}, V.~A. and {Macquart}, J.-P. and {Curran}, S.~J. and {Duchesne}, S.~W. and {Mahony}, E.~K. and {Sadler}, E.~M. and {Whiting}, M.~T. and {Bannister}, K.~W. and {Chippendale}, A.~P. and {Edwards}, P.~G. and {Harvey-Smith}, L. and {Heywood}, I. and {Indermuehle}, B.~T. and {Lenc}, E. and {Marvil}, J. and {McConnell}, D. and {Sault}, R.~J.},
        title = "{Illuminating the past 8 billion years of cold gas towards two gravitationally lensed quasars}",
      journal = {\mnras},
         year = 2017,
        month = mar,
       volume = {465},
       number = {4},
        pages = {4450-4467},
          doi = {10.1093/mnras/stw2860},
archivePrefix = {arXiv},
       eprint = {1611.00863},
 primaryClass = {astro-ph.GA},
       adsurl = {https://ui.adsabs.harvard.edu/abs/2017MNRAS.465.4450A}
}

@ARTICLE{Peroux2019,
       author = {{P{\'e}roux}, C{\'e}line and {Zwaan}, Martin A. and {Klitsch}, Anne and {Augustin}, Ramona and {Hamanowicz}, Aleksandra and {Rahmani}, Hadi and {Pettini}, Max and {Kulkarni}, Varsha and {Straka}, Lorrie A. and {Biggs}, Andy D. and {York}, Donald G. and {Milliard}, Bruno},
        title = "{Multiphase circumgalactic medium probed with MUSE and ALMA}",
      journal = {\mnras},
         year = 2019,
        month = may,
       volume = {485},
       number = {2},
        pages = {1595-1613},
          doi = {10.1093/mnras/stz202},
archivePrefix = {arXiv},
       eprint = {1901.05217},
 primaryClass = {astro-ph.GA},
       adsurl = {https://ui.adsabs.harvard.edu/abs/2019MNRAS.485.1595P}
}

@ARTICLE{Koay2011,
       author = {{Koay}, J.~Y. and {Bignall}, H.~E. and {Macquart}, J.-P. and {Jauncey}, D.~L. and {Rickett}, B.~J. and {Lovell}, J.~E.~J.},
        title = "{Detection of six rapidly scintillating active galactic nuclei and the diminished variability of J1819+3845}",
      journal = {\aap},
         year = 2011,
        month = oct,
       volume = {534},
          eid = {L1},
        pages = {L1},
          doi = {10.1051/0004-6361/201117805},
archivePrefix = {arXiv},
       eprint = {1109.1906},
 primaryClass = {astro-ph.GA},
       adsurl = {https://ui.adsabs.harvard.edu/abs/2011A&A...534L...1K}
}

@ARTICLE{Kochanek2012,
       author = {{Kochanek}, C.~S. and {Shappee}, B.~J. and {Stanek}, K.~Z. and {Holoien}, T.~W.-S. and {Thompson}, Todd A. and {Prieto}, J.~L. and {Dong}, Subo and {Shields}, J.~V. and {Will}, D. and {Britt}, C. and {Perzanowski}, D. and {Pojma{\'n}ski}, G.},
        title = "{The All-Sky Automated Survey for Supernovae (ASAS-SN) Light Curve Server v1.0}",
      journal = {\pasp},
         year = 2017,
        month = oct,
       volume = {129},
       number = {980},
        pages = {104502},
          doi = {10.1088/1538-3873/aa80d9},
archivePrefix = {arXiv},
       eprint = {1706.07060},
 primaryClass = {astro-ph.SR},
       adsurl = {https://ui.adsabs.harvard.edu/abs/2017PASP..129j4502K}
}

@article{Glowacki2019AnQuasars,
    title = {{An ASKAP survey for H i absorption towards dust-obscured quasars}},
    year = {2019},
    journal = {Monthly Notices of the Royal Astronomical Society},
    author = {Glowacki, M and Allison, J R and Moss, V A and Mahony, E K and Sadler, E M and Callingham, J R and Ellison, S L and Whiting, M T and Bunton, J D and Chippendale, A P and Heywood, I and McConnell, D and Raja, W and Voronkov, M A},
    number = {4},
    month = {11},
    pages = {4926--4943},
    volume = {489},
    doi = {10.1093/mnras/stz2452},
    issn = {0035-8711}
}

@article{Ross2015ExtremelySpectra,
    title = {{Extremely red quasars from SDSS, BOSS and <i>WISE</i> : classification of optical spectra}},
    year = {2015},
    journal = {Monthly Notices of the Royal Astronomical Society},
    author = {Ross, Nicholas P. and Hamann, Fred and Zakamska, Nadia L. and Richards, Gordon T. and Villforth, Carolin and Strauss, Michael A. and Greene, Jenny E. and Alexandroff, Rachael and Brandt, W. Niel and Liu, Guilin and Myers, Adam D. and P{\^{a}}ris, Isabelle and Schneider, Donald P.},
    number = {4},
    month = {11},
    pages = {3933--3953},
    volume = {453},
    doi = {10.1093/mnras/stv1710},
    issn = {0035-8711}
}

@article{Carilli1998RedshiftedQuasars,
    title = {{Redshifted Neutral Hydrogen 21 Centimeter Absorption toward Red Quasars}},
    year = {1998},
    journal = {The Astrophysical Journal},
    author = {Carilli, C. L. and Menten, Karl M. and Reid, Mark J. and Rupen, M. P. and Yun, Min Su},
    number = {1},
    month = {2},
    pages = {175--182},
    volume = {494},
    doi = {10.1086/305191},
    issn = {0004-637X}
}

@article{Dutta2020UGMRTQuasars,
    title = {{uGMRT search for cold gas at z ∼ 1–1.4 towards red quasars}},
    year = {2020},
    journal = {Monthly Notices of the Royal Astronomical Society},
    author = {Dutta, R and Raghunathan, S and Gupta, N and Joshi, R},
    number = {1},
    month = {1},
    pages = {838--847},
    volume = {491},
    doi = {10.1093/mnras/stz3084},
    issn = {0035-8711}
}

@article{Srianand2022EmergencePKS2355-106,
    title = {{Emergence of a new H  <scp>i</scp> 21-cm absorption component at <i>z</i> ∼ 1.1726 towards the <i>{$\gamma$}</i> -ray blazar PKS 2355-106}},
    year = {2022},
    journal = {Monthly Notices of the Royal Astronomical Society},
    author = {Srianand, Raghunathan and Gupta, Neeraj and Petitjean, Patrick and Momjian, Emmanuel and Balashev, Sergei A and Combes, Françoise and Chen, Hsiao-Wen and Krogager, Jens-Kristian and Noterdaeme, Pasquier and Rahmani, Hadi and Baker, Andrew J and Emig, Kimberly L and J{\'{o}}zsa, Gyula I G and Kloeckner, Hans-Rainer and Moodley, Kavilan},
    number = {1},
    month = {9},
    pages = {1339--1346},
    volume = {516},
    doi = {10.1093/mnras/stac1877},
    issn = {0035-8711}
}

@article{Gupta2018Revealing0.017,
    title = {{Revealing H i gas in emission and absorption on pc to kpc scales in a galaxy at z ∼ 0.017}},
    year = {2018},
    journal = {Monthly Notices of the Royal Astronomical Society},
    author = {Gupta, N and Srianand, R and Farnes, J S and Pidopryhora, Y and Vivek, M and Paragi, Z and Noterdaeme, P and Oosterloo, T and Petitjean, P},
    number = {2},
    month = {5},
    pages = {2432--2445},
    volume = {476},
    doi = {10.1093/mnras/sty384},
    issn = {0035-8711}
}

@article{Srianand2013Parsec-scale0.079,
    title = {{Parsec-scale structures and diffuse bands in a translucent interstellar medium at z≃ 0.079★}},
    year = {2013},
    journal = {Monthly Notices of the Royal Astronomical Society},
    author = {Srianand, R. and Gupta, N. and Rahmani, H. and Momjian, E. and Petitjean, P. and Noterdaeme, P.},
    number = {3},
    month = {1},
    pages = {2198--2206},
    volume = {428},
    doi = {10.1093/mnras/sts190},
    issn = {1365-2966}
}

@INPROCEEDINGS{Tody1993,
       author = {{Tody}, Doug},
        title = "{IRAF in the Nineties}",
    booktitle = {Astronomical Data Analysis Software and Systems II},
         year = 1993,
       editor = {{Hanisch}, R.~J. and {Brissenden}, R.~J.~V. and {Barnes}, J.},
       series = {Astronomical Society of the Pacific Conference Series},
       volume = {52},
        month = jan,
        pages = {173},
       adsurl = {https://ui.adsabs.harvard.edu/abs/1993ASPC...52..173T}
}

@ARTICLE{Beaklini2017,
       author = {{Beaklini}, Pedro P.~B. and {Dominici}, T{\^a}nia P. and {Abraham}, Zulema},
        title = "{Multiwavelength flaring activity of PKS 1510-089}",
      journal = {\aap},
         year = 2017,
        month = oct,
       volume = {606},
          eid = {A87},
        pages = {A87},
          doi = {10.1051/0004-6361/201731118},
archivePrefix = {arXiv},
       eprint = {1707.05795},
 primaryClass = {astro-ph.HE},
       adsurl = {https://ui.adsabs.harvard.edu/abs/2017A&A...606A..87B}
}


\end{document}